\documentclass[submission,copyright]{eptcs}
\providecommand{\event}{QPL 2014} 

\usepackage[T1]{fontenc}
\usepackage{textcomp}
\usepackage[utf8]{inputenc}

\usepackage{graphicx}
\usepackage{latexsym}
\usepackage{amssymb}
\usepackage{mathtools} 
\usepackage[amsmath,thmmarks,thref,hyperref]{ntheorem}
\usepackage{stmaryrd}
\usepackage{cmll}
\usepackage{mathrsfs}
\usepackage{dsfont}
\usepackage[noadjust]{cite}
\usepackage{bussproofs}
\usepackage{wrapfig}
\usepackage{enumitem}
\usepackage{array}

\allowdisplaybreaks[1] 

\usepackage[all,2cell]{xy}
\UseAllTwocells
\SelectTips{cm}{}  
\newdir{ >}{{}*!/-8pt/@{>}}
\newdir{|>}{%
!/1.6pt/@{|}*:(1,-.2)@^{>}*:(1,+.2)@_{>}}
\newdir{pb}{:(1,-1)@^{|-}}
  \def\pb#1{\save[]+<20 pt,0 pt>:a(#1)\ar@{pb{}}[]\restore}

\makeatletter
\newtheoremstyle{myplain}%
  {\item[\hskip\labelsep \theorem@headerfont ##1\ ##2\theorem@separator]}%
  {\item[\hskip\labelsep \theorem@headerfont ##1\ ##2\ %
    {\normalfont\unboldmath(##3)}\theorem@separator]}
\newtheoremstyle{nonumbermyplain}%
  {\item[\hskip\labelsep \theorem@headerfont ##1\theorem@separator]}%
  {\item[\hskip\labelsep \theorem@headerfont ##1\ %
    {\normalfont\unboldmath(##3)}\theorem@separator]}
\newtheoremstyle{myproof}%
  {\item[\hskip\labelsep \theorem@headerfont ##1\theorem@separator]}%
  {\item[\hskip\labelsep \theorem@headerfont ##3\theorem@separator]}
\makeatother

\theoremstyle{myplain}
\theoremheaderfont{\normalfont\bfseries}
\theorembodyfont{\itshape}
\theoremseparator{.}
\theoremsymbol{}

\newtheorem{mytheorem}{Theorem}[section]
\newtheorem{mylemma}[mytheorem]{Lemma}
\newtheorem{myproposition}[mytheorem]{Proposition}
\newtheorem{mycorollary}[mytheorem]{Corollary}

\theorembodyfont{\upshape}
\newtheorem{mydefinition}[mytheorem]{Definition}

\newtheorem{myremark}[mytheorem]{Remark}

\theoremheaderfont{\normalfont\itshape}

\theoremstyle{myproof}
\theoremsymbol{\ensuremath{\blacksquare}}
\newtheorem{myproof}{Proof}

\qedsymbol{\ensuremath{\blacksquare}}

\mathchardef\mhyphen="2D

\newcommand{\longto}{\longrightarrow}

\newcommand{\C}{\mathbb{C}}
\newcommand{\R}{\mathbb{R}}
\newcommand{\N}{\mathbb{N}}

\newcommand{\Set}{\mathbf{Set}}

\newcommand{\Cstar}{\mathbf{Cstar}}
\newcommand{\Wstar}{\mathbf{Wstar}}

\newcommand{\Dcppo}{\mathbf{Dcppo}}

\newcommand{\Dcppobot}{\Dcppo_\bot}

\newcommand{\omcppo}{\boldsymbol{\omega}\mathbf{Cppo}}

\newcommand{\CWstar}{\mathbf{CWstar}}
\newcommand{\FdWstar}{\mathbf{FdWstar}}

\newcommand{\CPM}{\mathbf{CPM}}
\newcommand{\CPMs}{\CPM_s}
\newcommand{\Q}{\mathbf{Q}}

\newcommand{\MI}{\mathrm{MI}}
\newcommand{\MIU}{\mathrm{MIU}}
\newcommand{\CP}{\mathrm{CP}}
\newcommand{\CPPU}{\mathrm{CPSU}}

\newcommand{\rmP}{\mathrm{P}}
\newcommand{\PPU}{\mathrm{PSU}}

\newcommand{\id}{\mathrm{id}}

\newcommand{\mat}{\mathcal{M}}
\newcommand{\spec}{\mathrm{Sp}}

\newcommand{\sa}[1]{#1_\mathrm{sa}}

\newcommand{\Ef}{\mathcal{E}\mspace{-4mu}f}

\newcommand{\calH}{\mathcal{H}}
\newcommand{\calK}{\mathcal{K}}
\newcommand{\calE}{\mathcal{E}}

\DeclareMathOperator*{\uwlim}{uw-lim}

\newcommand{\wtensor}{\mathop{\overline{\otimes}}}

\newcommand{\bsp}{\mathcal{B}}
\newcommand{\tcsp}{\mathcal{T}}

\newcommand{\conti}{\mathit{C}}

\newcommand{\op}{\mathrm{op}}
\newcommand{\bang}{\mathord{!}}

\DeclareMathOperator{\tr}{tr}
\DeclareMathOperator{\Tr}{Tr}
\DeclareMathOperator{\Fix}{Fix}

\newcommand{\Trn}[1]{\Tr^{(#1)}}
\newcommand{\Fixn}[1]{\Fix^{(#1)}}

\DeclarePairedDelimiter\abs{\lvert}{\rvert}
\DeclarePairedDelimiter\norm{\lVert}{\rVert}

\DeclarePairedDelimiter\tuple{\langle}{\rangle}

\DeclarePairedDelimiter\sem{\llbracket}{\rrbracket}

\newcommand{\lft}{\mathopen{}\mathclose\bgroup\left}
\newcommand{\rgt}{\aftergroup\egroup\right}
\newcommand{\mdl}{\mathrel{}\aftergroup\mathrel\aftergroup{\aftergroup}\middle}

\newcommand{\cat}[1]{\mathbf{#1}}

\newcommand{\catC}{\cat{C}}
\newcommand{\catD}{\cat{D}}

\newcommand{\longhookrightarrow}{\lhook\joinrel\longrightarrow}

\makeatletter
\newcommand{\verticalarrow}[4]{
\begingroup
\if\relax\detokenize{#3}\relax 
  \hbox{${{{\scriptstyle #2}\@@atop
  {\scriptstyle #1}}\@@atop
  {\scriptstyle #4}}$}
\else
  \setbox0=\hbox{${{{\scriptstyle #2}\@@atop
    {\scriptstyle #1}}\@@atop
    {\scriptstyle #4}}$}
  \setbox1=\hbox{$\scriptstyle #1 #3$}
  \setbox2=\hbox{$\scriptstyle #1$}
  \dimen0=.5\wd0
  \advance \dimen0 by \wd1
  \advance \dimen0 by -.5\wd2
  \ifdim\dimen0<\wd0 \dimen0=\wd0 \fi
  \hbox to \dimen0{${{{\scriptstyle #2}\@@atop
    {\scriptstyle #1 \mathrlap{#3}}}\@@atop
    {\scriptstyle #4}}$\hfill}
\fi
\endgroup
}
\newcommand{\fibop}[3]{
\begingroup
\setbox0=\hbox{${{{\scriptstyle #1}\@@atop
  {\scriptstyle \downarrow}}\@@atop
  {\scriptstyle #3}}$}
\setbox1=\hbox{$\scriptstyle #1^{(\op)}$}
\setbox2=\hbox{$\scriptstyle #1$}
\dimen0=\wd1
\advance \dimen0 by -.5\wd2
\if\relax\detokenize{#2}\relax 
  \advance \dimen0 by .5\wd0
  \ifdim\dimen0<\wd0 \dimen0=\wd0 \fi
  \hbox to \dimen0{${{{\scriptstyle #1^{\mathrlap{(\op)}}}\@@atop
    {\scriptstyle \downarrow}}\@@atop
    {\scriptstyle #3}}$\hfill}
\else
  \setbox3=\hbox{$\scriptstyle \downarrow #2^{(\op)}$}
  \setbox4=\hbox{$\scriptstyle \downarrow$}
  \dimen1=\wd3
  \advance \dimen1 by -.5\wd4
  \ifdim\dimen1>\dimen0 \dimen0=\dimen1 \fi
  \advance \dimen0 by .5\wd0
  \ifdim\dimen0<\wd0 \dimen0=\wd0 \fi
  \hbox to \dimen0{${{{\scriptstyle #1^{\mathrlap{(\op)}}}\@@atop
    {\scriptstyle \downarrow \mathrlap{#2^{(\op)}}}}\@@atop
    {\scriptstyle #3}}$\hfill}
\fi
\endgroup
}
\makeatother

\newif\ifignore 
\ignoretrue
\newcommand{\auxproof}[1]{
\ifignore\mbox{}\newline
\textbf{BEGIN: AUX-PROOF} \dotfill\newline
{\footnotesize #1}\mbox{}\newline
\textbf{END: AUX-PROOF}\dotfill\newline
\fi}

\newcommand{\noic}{\sb{}\kern-\scriptspace }

\DeclareMathAlphabet{\mathcal}{OMS}{cmsy}{m}{n}

\ignorefalse

\title{Semantics for a Quantum Programming Language\\
by Operator Algebras}
\author{Kenta Cho
\institute{Institute for Computing and Information Sciences (iCIS)\\
Radboud University Nijmegen, The Netherlands}
\email{%
\href{mailto:K.Cho@cs.ru.nl}{\nolinkurl{K.Cho@cs.ru.nl}}%
\textrm{,\enspace}%
\url{http://www.cs.ru.nl/K.Cho/}}
}
\def\titlerunning{Semantics for a Quantum Programming Language by Operator Algebras}
\def\authorrunning{K.~Cho}
\begin{document}
\maketitle

\begin{abstract}
This paper presents a novel semantics
for a quantum programming language
by \emph{operator algebras},
which are known to
give a formulation for quantum theory
that is alternative to
the one by Hilbert spaces.
We show that
the opposite category of
the category of $W^*$-algebras and normal completely positive subunital maps
is an elementary quantum flow chart category in the sense of Selinger.
As a consequence, it gives a denotational semantics for
Selinger's first-order functional
quantum programming language QPL.
The use of operator algebras allows us
to accommodate infinite structures and
to handle classical and quantum computations
in a unified way.
\end{abstract}

\section{Introduction}

Aiming at high-level and structured
description of quantum computation/information,
many \emph{quantum programming languages}
have been proposed and their semantics
studied~\cite{Gay2006,Valiron2013}.
As one of pioneering works,
in 2004 Selinger~\cite{Selinger2004} proposed
a first-order functional quantum programming language
QPL (or QFC),
and gave its denotational semantics
rigorously in terms of categories.
He (jointly with Valiron)
successively
started to study
a higher-order quantum programming language, or
the \emph{quantum lambda calculus}~\cite{SelingerV2006,SelingerV2008,SelingerV2009}.
It turned out to be challenging
to give a denotational semantics for the quantum lambda calculus
(with the full features, such as the $\oc$ modality and recursion).
The first denotational semantics
was given via Geometry of Interaction~\cite{HasuoH2011};
but there are
several different approaches~\cite{MalherbeSS2013,PaganiSV2014}.
As is stated in \cite[\S1]{PaganiSV2014},
the problem lies in
that quantum computation is typically modelled by
using finite dimensional Hilbert spaces,
and hence it is difficult to model
\emph{infinite} structures in computation.

The present paper proposes a novel denotational semantics for
a quantum programming language by \emph{operator algebras}.
Operator algebras, specifically $C^*$-algebras and $W^*$-algebras
(the latter are also known as von Neumann algebras),
give an alternative formulation for quantum theory
(sometimes called the \emph{algebraic} formulation~\cite{Landsman2009}).
It is worth mentioning that
von Neumann himself, who formulated quantum theory
by Hilbert spaces~\cite{Neumann1955},
developed the theory of operator
algebras~\cite{MurrayN1936,MurrayN1937,
Neumann1940,MurrayN1943,Neumann1949}
(some of them jointly with Murray),
and later preferred the algebraic approach
for quantum theory~\cite{Redei1996}.
Operator algebras have been successfully used
in areas such as
quantum statistical mechanics~\cite{BratteliR1987all2dois}
and quantum field theory~\cite{HaagK1964,Haag1996,Araki1999}.
They have also been of growing importance in
the area of quantum information~\cite{Keyl2002};
for example, in~\cite{DArianoKSW2007},
the impossibility of quantum bit commitment is (re)examined
in the algebraic formalism.

\subsection*{Contributions and related work}
In this paper it is shown that
the category $\Wstar_\CPPU$ of $W^*$-algebras and normal
completely positive subunital maps is
a $\Dcppobot$-enriched symmetric monoidal category
with $\Dcppobot$-enriched finite products.
It follows that its opposite
$(\Wstar_\CPPU)^\op$ is
an $\omcppo$-enriched
\emph{elementary quantum flow chart category}.
As a consequence, it gives rise to a denotational semantics
for a first-order functional quantum programming language QPL
designed by Selinger~\cite{Selinger2004}.

Selinger himself gave
a denotational semantics for QPL
by the category $\Q$~\cite{Selinger2004}.
In comparison to his original semantics,
our semantics by operator algebras has the following
two advantages.
First, our semantics accommodates infinite structures
such as a type $\mathbf{nat}$ of natural numbers.
This is because we discuss general $W^*$-algebras
and do not restrict them to finite dimensional ones.
In \S\ref{sect:embedding} we will see
that our model can be considered as an infinite dimensional
extension of Selinger's model.
Second, it is known that
\emph{commutative} $C^*$-algebras (resp.\ $W^*$-algebras)
correspond to and certain type of topological
(resp.\ measure) spaces,
and therefore we can interpret classical computation
as a map between commutative algebras.
This correspondence (or \emph{duality}) is studied well
for $C^*$-algebras e.g.\ in~\cite{FurberJ2013},
but not yet very well for $W^*$-algebras.
In this paper, hence,
we restrict it to deterministic computation
(in the category $\Set$),
and give rudimentary results
in \S\ref{sect:infinite-type}
as a first step.
It will enable us to handle
classical and quantum computations
in a unified way.
Although the present paper has not yet
been able to give
a semantics for a \emph{higher-order}
quantum programming language,
compared to such works~\cite{HasuoH2011,MalherbeSS2013,PaganiSV2014},
our semantics is quite simple:
a type is interpreted just as
a $W^*$-algebra, and a program as
a map of them (in opposite direction).
However, the theory of operator algebras itself
could be complicated enough.

As is the case in previous
works~\cite{Selinger2004,HasuoH2011,MalherbeSS2013,PaganiSV2014},
quantum computation is usually
modelled by using finite dimensional Hilbert spaces
$\C^n$ (or matrix algebras $\mat_n\cong\bsp(\C^n)$).
It seems that the (explicit) use of operator
algebras is not so common
in the area of quantum computation.
Recently there are several
works using $C^*$-algebras~\cite{FurberJ2013,Jacobs2013},
which led the author to the present work.
The use of $W^*$-algebras in this context
appeared independently and coincidentally
in Rennela's thesis~\cite{Rennela2013Master}
and the present work
(or the author's thesis~\cite{Cho2014Master}).
In~\cite[Theorem~3.8]{Rennela2013Master},
he also observed
that the category $\Wstar_\PPU$
of $W^*$-algebras and normal positive subunital maps
are $\Dcppo$-enriched,
which is almost the same result as
Theorem~\ref{thm:wstarcppu-dccpobot-enriched}
in the present paper.
In his latest paper~\cite{Rennela2014},
he further showed that
$\Wstar_\PPU$ is algebraically compact for
a certain class of functors.
This result enables us to have
inductively defined types.

Some related technical results
have appeared in~\cite{ChiribellaTU2013},
which studies spaces of maps (`quantum operations')
between $W^*$-algebras,
and maps (`quantum supermaps') between them.
For instance,
\cite[Proposition~7]{ChiribellaTU2013}
states (in our terms)
that $\Wstar_\CP(M,N)$ is bounded directed complete.

\subsection*{Organisation of the paper}
First we review complete partial orders
in \S\ref{sec:cpo},
and then review Selinger's work on QPL
in \S\ref{sec:QPL}.
In \S\ref{sec:opalg} we also review the basics
of operator algebras, and then look at
the order-theoretic perspective of operator algebras.
We show in \S\ref{sec:dcpo-enrichment}
that $\Wstar_\CPPU$ is
a $\Dcppobot$-enriched symmetric monoidal category with
$\Dcppobot$-enriched finite products
and in \S\ref{sec:sem-by-wstar} that
its opposite $(\Wstar_\CPPU)^\op$
is an $\omcppo$-enriched elementary quantum flow chart category,
hence gives a denotational semantics for QPL.
Section~\ref{sec:Schr-vs-Heis}
discuss a duality between Selinger's and our semantics,
and in \S\ref{sect:embedding} it is shown that Selinger's model
is contravariantly embedded into our model.
In \S\ref{sect:infinite-type} we see that our model can accommodate infinite types,
and also classical computation as commutative structures.
Section~\ref{sec:concl-future} concludes the paper with future work.

This paper is based on the author's master
thesis~\cite{Cho2014Master},
in which one may find further details.

\section{Complete partial orders}
\label{sec:cpo}

In this section, we will briefly review the notion
of complete partial orders,
which is essential in domain theory~\cite{AbramskyJ1994},
and is fundamental for
the denotational semantics of
programming languages.

\begin{mydefinition}
A poset is
\begin{enumerate}
\item
\emph{directed complete} if
every directed subset has a supremum;
\item
\emph{bounded directed complete} if
every directed subset
that is bounded from above
has a supremum;
\item
\emph{$\omega$-complete} if
every $\omega$-chain
($(x_n)_{n\in\omega}$ with $x_n\le x_{n+1}$)
has a supremum;
\item
\emph{pointed} if it has a least element (denoted by $\bot$).
\end{enumerate}
A (bounded) directed complete poset
is abbreviated as a \emph{(b)dcpo},
and an $\omega$-complete poset
as a \emph{$\omega$cpo}.
\end{mydefinition}

\begin{mydefinition}
A map between posets (in \ref{def-enum:strict}, pointed posets) is
\begin{enumerate}
\item
\emph{Scott-continuous}
if it preserves suprema of directed subsets;
\item
\emph{$\omega$-(Scott-)continuous}
if it preserves suprema of $\omega$-chains;
\item\label{def-enum:strict}
\emph{strict} if it preserves the least element.
\end{enumerate}
\end{mydefinition}
Note that every dcpo is an $\omega$cpo,
and every Scott-continuous map is $\omega$-continuous.
The next theorem is a fundamental tool
to give an interpretation of a program with recursion and loop.
\begin{mytheorem}
Every $\omega$-continuous endomap $f$
on a pointed $\omega$cpo has a least fixed point,
which is given by $\bigvee_n f^n(\bot)$.
\qed
\end{mytheorem}

We here fix the notations of categories we use
in this paper.
\begin{mydefinition}
We denote by
$\Dcppobot$
the category of
pointed dcpos and strict Scott-continuous maps,
and by
$\omcppo$ the category of
pointed $\omega$cpos and $\omega$-continuous maps.
\end{mydefinition}
Both categories $\Dcppobot$ and $\omcppo$
have products, which are given by
cartesian products of underlying sets with
coordinatewise order.
When we speak of $\Dcppobot$- (or $\omcppo$)-enrichment
of categories,
these cartesian structures are taken
as monoidal structures.

\section{Selinger's QPL and its semantics}
\label{sec:QPL}

In \cite{Selinger2004},
Selinger proposed
a quantum programming language
QPL and its denotational semantics.
It is a first-order functional language
with loop and recursion,
and is described by
flow chart syntax, besides
by textual syntax.\footnote{%
Strictly speaking,
the name QPL is reserved
for the textual syntax.
In the present paper
we do not distinguish the two syntaxes
because semantics are given in the same manner.}
He showed a denotational semantics
for QPL is given by the following category.
\begin{mydefinition}[{\cite[\S6.6]{Selinger2004}}]
\label{def:elem-quantum-flow-chart-cat}
An \emph{elementary quantum flow chart category}
is a symmetric monoidal category
$(\catC, \otimes, I)$ with
traced finite coproducts
$(\oplus, 0, \Tr)$
such that:
\begin{itemize}
\item
For each $A\in\catC$,
$A\otimes (-)$ is a traced monoidal functor.
\item
$\catC$ has a distinguished object $\mathbf{qbit}$
with arrows $\iota\colon I\oplus I\to\mathbf{qbit}$
and $p\colon \mathbf{qbit}\to I\oplus I$
such that $p\circ \iota = \id$.
\end{itemize}
\end{mydefinition}

\begin{mytheorem}[{\cite[\S6.6]{Selinger2004}}]
\label{thm:eqfcc-give-sem-qpl}
Let $\catC$ be an
elementary quantum flow chart category.
Suppose we have an assignment $\eta$ of
built-in unitary operator symbols $S$
of arity $n$
to arrows
$\eta_S\colon\mathbf{qbit}^{\otimes n}\to
\mathbf{qbit}^{\otimes n}$ in $\catC$.
Then we have an interpretation of QPL programs
without recursion.
If $\catC$ is additionally $\omcppo$-enriched,
then we can also interpret QPL programs with recursion.
\qed
\end{mytheorem}
Selinger also gave a concrete model
for QPL,
constructing
an $\omcppo$-enriched elementary quantum flow chart category
$\Q$ as follows.

\begin{mydefinition}
\label{def:selingers-cat}
We write $\mat_n$
for the set of complex $n\times n$
matrices.
\begin{enumerate}
\item
The category $\CPMs$
is defined as follows.
\begin{itemize}
\item
An object is a natural number.
\item
An arrow $f\colon n\to m$ is a
completely positive map
$f\colon \mat_n\to \mat_m$.
\end{itemize}
\item
The category $\CPM$
is the finite biproduct completion
of $\CPMs$.
Specifically:
\begin{itemize}
\item
An object is
a sequence $\vec{n}=(n_1,\dotsc,n_k)$ of natural numbers.
\item
An arrow $f\colon \vec{n}\to \vec{m}$
is a matrix $(f_{ij})$
of arrows $f_{ij} \colon n_j\to m_i$ in $\CPMs$.
\end{itemize}
\item
\label{def:cat-Q}
The category $\Q$
is a subcategory of $\CPM$
such that
\begin{itemize}
\item
Objects are the same as
$\CPM$.
\item
An arrow is $f\colon \vec{n}\to \vec{m}$
in $\CPM$ which is trace-nonincreasing, i.e.
\[
\sum\nolimits_i
\sum\nolimits_j
\tr(f_{ij}(A_j))
\le
\sum\nolimits_j\tr(A_j)
\]
for all $(A_j)_j$ with positive $A_j\in\mat_{n_j}$.
\end{itemize}
\end{enumerate}
\end{mydefinition}

The category $\Q$
turns out to have the monoidal structure
$(\otimes, I)$ and
finite coproducts $(\oplus, 0)$,
and furthermore to be $\omcppo$-enriched.
The category is also equipped with
a monoidal trace w.r.t.\ $(\oplus, 0)$,
which
is obtained from its $\omcppo$-enriched structure.
Then it forms
an $\omcppo$-enriched elementary quantum flow chart category
with $\mathbf{qbit}\coloneqq 2$.
As Selinger mentioned~\cite[\S6.4]{Selinger2004}
(though he did not give a proof),
the construction of monoidal trace
from $\omcppo$-enriched structure
works for every $\omcppo$-enriched category
with finite coproducts
(satisfying certain condition).
Specifically, we have the following theorem.
\begin{mytheorem}
\label{thm:cppo-enriched-cart-cat}
\mbox{}
\begin{enumerate}
\item\label{thm-enum:cppo-enriched-cart-cat}
Every $\omcppo$-enriched cocartesian category
with right-strict composition
(i.e.\ $f\circ\bot=\bot$)
is traced.

\item
Let $\catC$ and $\catD$
be $\omcppo$-enriched cocartesian categories
with right-strict composition,
which are traced by
\ref{thm-enum:cppo-enriched-cart-cat}.
Every
$\omcppo$-enriched cocartesian functor
between $\catC$ and $\catD$
satisfying $F\bot=\bot$
is traced.
\end{enumerate}
\qed
\end{mytheorem}
Here,
a cocartesian category refers
to a monoidal category whose monoidal
structure is given by finite coproducts.
The proof is found in
Appendix~\ref{sec:trace-on-omcppo-enrieched-cat}.
Combining
Theorem~\ref{thm:cppo-enriched-cart-cat}
with
Definition~\ref{def:elem-quantum-flow-chart-cat},
we obtain a sufficient condition
for a category to be
an $\omcppo$-enriched elementary quantum flow chart category,
which gives a semantics for QPL programs
with recursion.
\begin{mytheorem}
\label{thm:omcppo-eqfcc}
A category is
an $\omcppo$-enriched elementary quantum flow chart category
if it is
an $\omcppo$-enriched symmetric monoidal category
$(\catC, \otimes, I)$
satisfying the following conditions.
\begin{itemize}
\item
$\catC$ has $\omcppo$-enriched finite coproducts
$(\oplus, 0)$;
\item
the composition is right-strict;
\item
for each $A\in\catC$,
a functor $A\otimes (-)$
preserves finite coproducts
and bottom arrows;
\item
$\catC$ has a distinguished object $\mathbf{qbit}$
with arrows $\iota\colon I\oplus I\to\mathbf{qbit}$
and $p\colon \mathbf{qbit}\to I\oplus I$
such that $p\circ \iota = \id$.
\end{itemize}
\qed
\end{mytheorem}

\section{Operator algebras: $C^*$-algebras and $W^*$-algebras}
\label{sec:opalg}

\subsection{$C^*$-algebras and $W^*$-algebras}

The theory of operator algebras is usually concerned
with $C^*$-algebras and $W^*$-algebras
(the latter are often studied as von Neumann algebras).
Due to limitations of space,
here we just fix notations and terminology,
and list the basic results.
A reader who is not familiar with these topics may
consult \cite{Takesaki2002,Sakai1998}
for the standard theory
and \cite{Meyer2008,Kornell2012} for the categorical perspective.

In this paper, $C^*$-algebras are always
assumed to be unital.
Note also that a `map' usually refers to
a \emph{linear} map.
For a $C^*$-algebra $A$,
we write $\sa{A}$ for
the set of self-adjoint elements,
and $[0,1]_A\coloneqq \{x\in A\mid 0\le x\le 1\}$
for the ``unit interval'',
or the set of \emph{effects}.
We write $\Cstar$ for the category of
$C^*$-algebras and linear maps.
We will denote subcategories of $\Cstar$,
with the same objects but different maps,
by adding subscripts to $\Cstar$ as follows:
M for `multiplicative'; I for `involutive';
P for `positive'; CP for `completely positive';
U for `unital'; SU for `subunital'\footnote{%
A map $f$ between $C^*$-algebras is said to be \emph{subunital}
if $f(1)\le 1$.
Note that in the author's thesis~\cite{Cho2014Master}
(and the preliminary version of this paper),
`pre-unital' is used instead of `subunital'.}.
For example,
$\Cstar_\MIU$ is the category of
$C^*$-algebras and unital $*$-homomorphisms
(i.e.\ multiplicative involutive maps),
while $\Cstar_\CPPU$ is the category of
$C^*$-algebras and completely positive subunital maps.
Note that there are inclusions like
$\Cstar_\MI\subseteq\Cstar_\CP\subseteq\Cstar_\rmP$.

A \emph{$W^*$-algebra} is a $C^*$-algebra that has
a (necessarily unique) predual.
Every $W^*$-algebra is equipped with the weak* topology
introduced by its predual, which is called the
\emph{ultraweak} topology. A map between $W^*$-algebras
is said to be \emph{normal} if it is ultraweakly continuous.
We write $\Wstar$ for the category of
$W^*$-algebras and normal maps.
We denote subcategories of $\Wstar$ in the same manner as $\Cstar$.
For example, $\Wstar_\MIU$ is
the category of $W^*$-algebras and normal unital $*$-homomorphisms,
which is a \emph{non-full} subcategory of $\Cstar_\MIU$
(since maps in $\Wstar$ are required to be normal).

We denote
\emph{direct sum}~\cite[Definition~1.1.5]{Sakai1998}
of $C^*$- and $W^*$-algebras by $\oplus$.
It forms categorical products
in categories $\Cstar_\MIU$,
$\Cstar_\CP$, $\Cstar_\CPPU$,
$\Wstar_\MIU$,
$\Wstar_\CP$, $\Wstar_\CPPU$ etc.
The nullary
direct sum is the zero space $0$.

There are several kinds of tensor products
for $C^*$- and $W^*$-algebras.
In this paper we use the \emph{spatial $C^*$-tensor
product}~\cite[Definition~1.22.8]{Sakai1998},
denoted by $\otimes$,
for $C^*$-algebras,
and the \emph{spatial $W^*$-tensor
product}~\cite[Definition~1.22.10]{Sakai1998},
denoted by $\wtensor$,
for $W^*$-algebras.
The spatial $C^*$-tensor product
makes categories $\Cstar_\MIU$,
$\Cstar_\CP$, $\Cstar_\CPPU$ etc.\ symmetric monoidal categories
with the unit object $\C$,
but not $\Cstar_\rmP$. In fact, a tensor product of
positive maps can be unbounded~\cite[Proposition~3.5.2]{BrownO2008}.
This is why we need the notion of the \emph{complete} positivity.
Similarly,
the spatial $W^*$-tensor
product makes categories $\Wstar_\MIU$,
$\Wstar_\CP$, $\Wstar_\CPPU$ etc.\ symmetric monoidal categories.

The spatial $C^*$- and $W^*$-tensor products
distribute over finite direct sums,
i.e.\
\[
A\otimes (B\oplus C)\cong (A\otimes B)\oplus (A\otimes C)
\enspace,\quad
A\otimes 0\cong 0
\enspace,\quad
M\wtensor (N\oplus L)\cong (M\wtensor N)\oplus (M\wtensor L)
\enspace,\quad
M\wtensor 0\cong 0
\enspace,
\]
for $C^*$-algebras $A,B,C$ and $W^*$-algebras $M,N,L$.
These properties seem to be known results, but
to be missing in the standard literature.
For the sake of completeness,
the proofs are included in Appendix~\ref{sec:dist-tensorprod-directsum}.
For our purpose,
it is useful to restate them as follows.

\begin{myproposition}
\label{prop:tensor-distr}
\mbox{}
\begin{enumerate}
\item
For each $C^*$-algebra $A$,
a functor $A\otimes(-)$ on $\Cstar_\MIU$
preserves finite products.
\item\label{prop:enum:wtensor-distr}
For each $W^*$-algebra $M$,
a functor $M\wtensor(-)$ on $\Wstar_\MIU$
preserves finite products.
\end{enumerate}
\qed
\end{myproposition}

\subsection{Order theory in operator algebras}

Recall each $C^*$-algebra is equipped with
a partial order $\le$ defined by: $a\le b$ $\Longleftrightarrow$
`$b-a$ is positive'.
Many concepts on operator algebras can be rephrased
in terms of the orders.
Observe, for instance, the following easy proposition.

\begin{myproposition}
A map between $C^*$-algebras is positive if and only if
it is monotone.
\qed
\end{myproposition}

In fact,
the orders on $W^*$-algebras have a
very nice property,
called \emph{monotone completeness},
which distinguishes $W^*$-algebras from
$C^*$-algebras.

\begin{mydefinition}
A $C^*$-algebra $A$ is \emph{monotone complete}
(or \emph{monotone closed})
if every norm-bounded directed subset of $\sa{A}$ has
the supremum in $\sa{A}$.
\end{mydefinition}

\begin{myproposition}[{\cite[Lemma~1.7.4]{Sakai1998}}]
\label{prop:wstar-monotone-complete}
Every $W^*$-algebra is monotone complete.
Moreover, suprema
are obtained as ultraweak limits.
\qed
\end{myproposition}

\noindent
We can also rephrase
the notion of normality of positive (i.e.\ monotone) maps
between $W^*$-algebras.
\begin{myproposition}[{\cite[Corollary~46.5]{Conway2000}}]
\label{prop:normal-sup-pres-prelim}
Let $f\colon M\to N$ be a positive map between
$W^*$-algebra. Then
$f$ is normal (i.e.\ ultraweakly continuous)
if and only if
it preserves the supremum of
every norm-bounded directed subset of $\sa{M}$.
\qed
\end{myproposition}

\noindent
Therefore, we shall say a positive map
$f\colon A\to B$
between monotone complete $C^*$-algebras
is \emph{normal} if
it preserves the supremum of
every norm-bounded directed subset of $\sa{A}$.
Then $W^*$-algebras can be characterised as follows.

\begin{mytheorem}[{\cite[Theorem~III.3.16]{Takesaki2002}}]
A $C^*$-algebra is a $W^*$-algebra if and only if
it is monotone complete and admits sufficiently
many normal positive functionals
(i.e.\ they separate the points).
\qed
\end{mytheorem}

Now, we shall recapture the order structures
in operator algebras
from a more
order-theoretic (or \emph{domain-theoretic}~\cite{AbramskyJ1994})
point of view.

\begin{myproposition}
\label{prop:equiv-monotone-closed}
Let $A$ be a $C^*$-algebra.
The following are equivalent.
\begin{enumerate}
\item\label{prop-enum:norm-bounded}
$A$ is monotone complete.
\item\label{prop-enum:order-bounded-above}
$\sa{A}$ is bounded directed complete.
\item\label{prop-enum:unit-interval}
$[0,1]_A$ is directed complete.
\end{enumerate}
\end{myproposition}
\begin{myproof}
Without loss of generality,
we may assume directed subsets are bounded from below.
Then,
\ref{prop-enum:norm-bounded} $\Longleftrightarrow$
\ref{prop-enum:order-bounded-above} follows from
the fact that
norm-boundedness and order-boundedness
coincide (see Proposition~\ref{prop:norm-bdd-vs-order-bdd}
in Appendix~\ref{sec-apx:other-omitted-proofs}).

\ref{prop-enum:order-bounded-above} $\Longrightarrow$
\ref{prop-enum:unit-interval}
is trivial.
For the converse,
note that $\sa{A}$ is an ordered vector space over $\R$.
Hence we can obtain the supremum
of a bounded directed subset of $\sa{A}$
by transforming it into
a directed subset of $[0,1]_A$ by shifting and scaling.
\end{myproof}

\noindent
Consequently, for every $W^*$-algebras $M$,
$\sa{M}$ is a bdcpo, and $[0,1]_M$ is a pointed dcpo.
We have a corresponding result for normal maps,
which is proved in a similar way,
using Proposition~\ref{prop:normal-sup-pres-prelim}.
\begin{myproposition}
\label{prop:normal-sup-pres}
Let $f\colon M\to N$ be a positive map between
$W^*$-algebra. The first two of the following are equivalent.
They are also equivalent to the third
when $f$ is subunital.
\begin{enumerate}
\item
$f$ is normal.
\item
The restriction
$f|_{\sa{M}}\colon\sa{M}\to\sa{N}$
is Scott-continuous.
\item
The restriction
$f|_{[0,1]_M}\colon [0,1]_M\to [0,1]_N$
is Scott-continuous.
\end{enumerate}
\qed
\end{myproposition}

\section{$\Dcppobot$-enrichment of the category of $W^*$-algebras}
\label{sec:dcpo-enrichment}

In this section we will show that
the category $\Wstar_\CPPU$ is $\Dcppobot$-enriched.
We also see that the monoidal product $(\wtensor, \C)$
and finite products $(\oplus, 0)$ on $\Wstar_\CPPU$
are $\Dcppobot$-enriched.

\begin{mydefinition}
\label{def:order-hom}
Let $M,N$ be $W^*$-algebras.
We define a partial order $\sqsubseteq$ on $\Wstar_\CPPU(M,N)$
by
\[
f\sqsubseteq g
\stackrel{\mathrm{def}}{\iff}
g-f \text{ is completely positive}
\enspace.
\]
\end{mydefinition}

\begin{myproposition}
\label{prop:hom-wstarcppu-is-dcppo}
For any $W^*$-algebras $M$ and $N$,
$\Wstar_\CPPU(M,N)$ with the order $\sqsubseteq$
is a pointed dcpo.
\end{myproposition}
\begin{myproof}
First of all,
it is easy to see that
the zero map is the least element of $\Wstar_\CPPU(M,N)$,
hence it is pointed.

Note that a positive subunital map
$f\colon A\to B$
between $C^*$-algebras restricts to $[0,1]_A\to [0,1]_B$,
while a ``linear''\footnote{%
$g(0)=0$, $g(x+y)=g(x)+g(y)$
for $x+y\le 1$,
$g(rx)=r g(x)$ for $r\in[0,1]$.}
map $g\colon [0,1]_A\to [0,1]_B$
extends to a positive subunital map
$A\to B$.
Hence such maps correspond bijectively
(cf.\ \cite[Lemma~2]{FurberJ2013}).

Let $(f_i)$ be a monotone net in $\Wstar_\CPPU(M,N)$.
For each $x\in[0,1]_M$,
$(f_i(x))$ is a monotone net in $[0,1]_N$,
which is a dcpo.
Hence define $f(x)\coloneqq\sup f_i(x)$.
It is easy to see $f$ is a ``linear'' map
$[0,1]_M\to[0,1]_N$,
so that we obtain a positive subunital map $f\colon M\to N$.

Normality of $f$ is proved as follows.
For a monotone net $(x_j)$
in $[0,1]_M$,
\[
f\Bigl(\sup x_j\Bigr)
= \sup_i f_i\Bigl(\sup_j x_j\Bigr)
= \sup_i \Bigl(\sup_j f_i(x_j)\Bigr)
= \sup_j \Bigl(\sup_i f_i(x_j)\Bigr)
= \sup_j f(x_j)
\enspace.
\]
Note that we can exchange of the order of $\sup$
(\cite[Proposition~2.1.12]{AbramskyJ1994}).
Therefore $f\colon[0,1]_M\to[0,1]_N$ is Scott-continuous,
and $f\colon M\to N$ is normal by
Proposition~\ref{prop:normal-sup-pres}.

To show that $f$ is completely positive
(hence $f\in\Wstar_\CPPU(M,N)$)
and that $f$ is indeed the supremum of $(f_i)_i$,
we need some more results on $W^*$-algebras.
The remaining proof is found at
Proposition~\ref{prop:hom-sup-cp} in
Appendix~\ref{sec-apx:other-omitted-proofs}.
\end{myproof}

This shows that
every hom-set of $\Wstar_\CPPU$
is an object in $\Dcppobot$.
We can also show that
the composition in $\Wstar_\CPPU$
is bi-strict Scott-continuous
(see Proposition~\ref{prop:composition-scott-conti}
in Appendix~\ref{sec-apx:other-omitted-proofs}).
Therefore, now we prove:

\begin{mytheorem}
\label{thm:wstarcppu-dccpobot-enriched}
The category $\Wstar_\CPPU$ is $\Dcppobot$-enriched.
Moreover, the composition is bi-strict,
i.e.\ $\bot\circ f=\bot$ and $f\circ\bot=\bot$
for each arrow $f$.
\qed
\end{mytheorem}

\noindent
Moreover, finite products $(\oplus,0)$ and
the monoidal product $(\otimes,0)$
are also suitably enriched:

\begin{mytheorem}
\label{thm:finprod-dcppobot-enriched}
Finite products in $\Wstar_\CPPU$ are
$\Dcppobot$-enriched.
\qed
\end{mytheorem}

\begin{mytheorem}
\label{thm:monoprod-dcppobot-enriched}
The symmetric monoidal product $(\wtensor,\C)$
on $\Wstar_\CPPU$ is
$\Dcppobot$-enriched.
Moreover, the tensor product of maps is bi-strict,
i.e.\ $\bot\wtensor f=\bot$ and $f\wtensor\bot=\bot$.
\qed
\end{mytheorem}

\noindent
The details are found in
Propositions~\ref{prop:tupling-scott-conti}
and~\ref{prop:wtensor-scott-conti}
in Appendix~\ref{sec-apx:other-omitted-proofs}, respectively.

\begin{myremark}
It is worth noting that the category $\Cstar_\CPPU$
is never $\Dcppo_\bot$-enriched, nor $\omcppo$-enriched.
This is because we have an order-isomorphism
$\Cstar_\CPPU(\C,A)\cong [0,1]_A$,
whereas there exists a $C^*$-algebra
such that $[0,1]_A$ is not $\omega$-complete
(take $A=\conti([0,1])$ for example).
\end{myremark}

\section{Semantics for QPL by $W^*$-algebras}
\label{sec:sem-by-wstar}

We have proved that
$\Wstar_\CPPU$ is an
$\Dcppobot$-enriched symmetric monoidal category with
$\Dcppobot$-enriched finite products.
Now we can show
the following theorem.

\begin{mytheorem}
The opposite category of $\Wstar_\CPPU$
is an $\omcppo$-enriched
elementary quantum flow chart category.
\end{mytheorem}
\begin{myproof}
We apply Theorem~\ref{thm:omcppo-eqfcc}.
All requirements follow from
Theorems~\ref{thm:wstarcppu-dccpobot-enriched},
\ref{thm:finprod-dcppobot-enriched},
\ref{thm:monoprod-dcppobot-enriched}
($\Dcppobot$-enrichment implies $\omcppo$-enrichment), and
Proposition~\ref{prop:tensor-distr}.\ref{prop:enum:wtensor-distr},
except a distinguished object $\mathbf{qbit}$
with arrows $\iota$, $p$.

We take $\mathbf{qbit}\coloneqq \mat_2$,
the algebra of complex $2\times 2$-matrices.
We define two maps $\iota$, $p$ by
\[
\iota
\left(
\begin{bmatrix}
x & y \\
z & w
\end{bmatrix}
\right)
= (x,w)
\enspace,\qquad\qquad
p(x,y) =
\begin{bmatrix}
x & 0 \\
0 & y
\end{bmatrix}
\enspace.
\]
It is straightforward to see the two maps are positive,
hence completely positive by
\cite[Corollary~IV.3.5 and Proposition~IV.3.9]{Takesaki2002}
(notice that $\C\oplus\C$ is commutative).
They are normal because they are maps between finite dimensional
$W^*$-algebras.
Moreover they are clearly unital.
Therefore $\iota$ and $p$ are arrows in $\Wstar_\CPPU$.
It is clear that $\iota\circ p = \id$, hence
$p\circ\iota = \id$ in $(\Wstar_\CPPU)^\op$.
\end{myproof}

Every unitary operator $S\colon (\C^2)^{\otimes n}\to(\C^2)^{\otimes n}$
of arity $n$ determines an arrow $\eta_S\colon
(\mat_2)^{\wtensor n}\to (\mat_2)^{\wtensor n}$ in $\Wstar_\CPPU$
by
\[
(\mat_2)^{\wtensor n}\cong \mat_{2^n}
\longto
\mat_{2^n}\cong
(\mat_2)^{\wtensor n}
\enspace,\qquad
x\longmapsto S^\dagger x S
\enspace,
\]
where $S$ is seen as a complex $2^n\times 2^n$ matrix.
By Theorem~\ref{thm:eqfcc-give-sem-qpl},
finally, we show:
\begin{mytheorem}
The opposite category of $\Wstar_\CPPU$
gives a denotational semantics for QPL (with recursion).
\qed
\end{mytheorem}

\section{Schr\"odinger vs.\ Heisenberg picture}
\label{sec:Schr-vs-Heis}

We have shown the category $(\Wstar_\CPPU)^\op$
gives a semantics for QPL.
From now on we will discuss a comparison
between two semantics by Selinger's $\Q$ and
our $(\Wstar_\CPPU)^\op$.

Recall that for a Hilbert space $\calH$,
$\bsp(\calH)$, i.e.\ the set of bounded operators on $\calH$,
is a $W^*$-algebra with the predual
$\tcsp(\calH)$, i.e.\ the set of trace class operators on $\calH$.
For every normal map
$\calE\colon\bsp(\calH)\to\bsp(\calK)$,
there exists a corresponding bounded map
$\calE_*\colon\tcsp(\calK)\to\tcsp(\calH)$
between preduals. They are
related in the following way:
\begin{equation}
\label{eq:duality}
\tr(\calE(S)\cdot T) = \tr(S\cdot \calE_*(T))
\end{equation}
for all $S\in\bsp(\calH)$ and $T\in\tcsp(\calK)$.
Furthermore the following holds.
\begin{myproposition}[{\cite[\S4.1.2]{HeinosaariZ2012}}]
Let $\calH$ and $\calK$ be Hilbert spaces.
Suppose
a normal map
$\calE\colon\bsp(\calH)\to\bsp(\calK)$
and
a bounded map
$\calE_*\colon\tcsp(\calK)\to\tcsp(\calH)$
related as above.
Then
\begin{enumerate}
\item
$\calE$ is completely positive
if and only if
$\calE_*$ is completely positive.
\item
$\calE$ is unital
if and only if
$\calE_*$ is trace-preserving.
\item
$\calE$ is subunital
if and only if
$\calE_*$ is trace-nonincreasing.
\end{enumerate}
\qed
\end{myproposition}

Hence, a normal completely positive subunital map
$\calE\colon\bsp(\calH)\to\bsp(\calK)$,
which is an arrow in $\Wstar_\CPPU$,
corresponds to
a completely positive trace-nonincreasing map
$\calE_*\colon\tcsp(\calK)\to\tcsp(\calH)$,
which is known as a \emph{quantum operation}
(see e.g.~\cite[\S8.2]{NielsenC2000},
\cite[Chap.~4]{HeinosaariZ2012}).
This is the well-known duality between
the Heisenberg and Schr\"odinger pictures:
one transforms observables
(i.e.\ self-adjoint operators),
while another transforms states
(i.e.\ density operators).

Hence, it is understood that
our semantics for QPL
by $\Wstar_\CPPU$
is given in the Heisenberg picture,
while Selinger's semantics
by $\Q$ is given in the Schr\"odinger picture.
In the words of
\cite{DHondtP2006},
our semantics
can also be thought of as the \emph{weakest precondition}
semantics.
This is because a positive subunital map
$\calE\colon\bsp(\calH)\to\bsp(\calK)$
can be restricted to a map
$\calE\colon\Ef(\calH)\to\Ef(\calK)$
between their effects,
where $\Ef(\calH)\coloneqq[0,1]_{\bsp(\calH)}$
is the set of effects on $\calH$,
and coincides with the set of \emph{predicates}
in \cite{DHondtP2006}.

\section{Embedding $\Q$ into $(\Wstar_\CPPU)^\op$}
\label{sect:embedding}

As seen in the previous section,
the two semantics by $\Wstar_\CPPU$ and $\Q$
can be considered as
different viewpoints (Schr\"odinger vs.\ Heisenberg)
for the same phenomena.
We can state it categorically:
the category $\Q$ can be
contravariantly embedded into $\Wstar_\CPPU$.

First we show the following embedding.
\begin{mytheorem}
There is a full embedding $I\colon\CPM\to(\Wstar_\CP)^\op$.
\qed
\end{mytheorem}
\begin{myproof}
Observe
the following bijective correspondences.
\[
\renewcommand{\arraystretch}{1.2}
\begin{array}{c}
f\colon (n_1,\dotsc,n_k)\longto
(m_1,\dotsc,m_l)
\quad\text{in $\CPM$}
\\ \hline\hline
f_{ij}\colon n_j\longto m_i
\quad\text{in $\CPMs$,
for each $i,j$}
\\ \hline\hline
f_{ij}\colon \mat_{n_j}\longto \mat_{m_i}
\quad\text{completely positive,
for each $i,j$}
\\ \hline\hline
(f_{ij})^*\colon \mat_{m_i}\longto \mat_{n_j}
\quad\text{completely positive,
hence an arrow in $\Wstar_\CP$,
for each $i,j$}
\\ \hline\hline
I (f)
 \colon \bigoplus_{i=1}^l\mat_{m_i}
\longto
\bigoplus_{j=1}^k\mat_{n_j}
\quad\text{in $\Wstar_\CP$}
\end{array}
\]
For the third correspondence,
note that the self-duality of
finite dimensional spaces:
\[
\mat_{n_j}
\cong
\bsp(\C^{n_j})
\cong
\tcsp(\C^{n_j})^*
\cong
(\mat_{n_j})^*
\enspace.
\]
The last correspondence comes from
the fact finite direct sums are biproducts
in $\Wstar_\CP$.
Hence the mapping
$I(n_1,\dotsc,n_k)=\bigoplus_{j=1}^k\mat_{n_j}$
defines a contravariant functor
$I\colon\CPM\to(\Wstar_\CP)^\op$,
which is full and faithful
by definition,
and clearly injective on objects.
\end{myproof}

We use the following lemma.

\begin{mylemma}
\label{lem:pos-trace}
Let $\calH$ be a Hilbert space.
A bounded operator $T\in\bsp(\calH)$
is positive if and only if
$\tr(TS)\in\R^+$ ($=[0,\infty)$) for all
positive $S\in\tcsp(\calH)$.
\qed
\end{mylemma}

\begin{mytheorem}
There is a full embedding $I'\colon\Q\to(\Wstar_\CPPU)^\op$.
\end{mytheorem}
\begin{myproof}
The functor
$I\colon\CPM\to(\Wstar_\CP)^\op$
restricts to a full and faithful functor
$I'\colon\Q\to(\Wstar_\CPPU)^\op$
as follows.
\begin{align*}
&
\text{$f\colon \vec{n}\to \vec{m}$
is trace-nonincreasing
(Definition~\ref{def:selingers-cat}.\ref{def:cat-Q})}
\\
&\iff
\sum\nolimits_i
\sum\nolimits_j
\tr(f_{ij}(A_j))
\le
\sum\nolimits_j\tr(A_j)
\text{ for all $(A_j)_j$ with positive $A_j\in \mat_{n_j}$}
\\
&\iff
\sum\nolimits_i
\tr(f_{ij}(A))
\le
\tr(A)
\text{ for each $A\in \mat_{n_j}$, for each $j$}
\\
&\stackrel{\star}{\iff}
\sum\nolimits_i
\tr\Bigl(\bigl((f_{ij})^*(1)\bigr)A\Bigr)
\le
\tr(A)
\text{ for each $A\in \mat_{n_j}$, for each $j$}
\\
&\iff
\tr\Bigl(
\bigl(1 -
\sum\nolimits_i
(f_{ij})^*(1)\bigr)A\Bigr)
\ge 0
\text{ for each $A\in \mat_{n_j}$, for each $j$}
\\
&\stackrel{\star\star}{\iff}
1 -
\sum\nolimits_i
(f_{ij})^*(1)
\ge 0
\text{ for each $j$}
\\
&\iff
\sum\nolimits_i
(f_{ij})^*(1)
\le 1
\text{ for each $j$}
\\
&\iff
I(f)\bigl((1)_i\bigr) \le
(1)_j
\\
&\iff
I(f)\colon \bigoplus\nolimits_i\mat_{m_i}
\longto
\bigoplus\nolimits_j\mat_{n_j}
\text{ is subunital}
\enspace,
\end{align*}
where $\stackrel{\star}{\Longleftrightarrow}$
is by the equation \eqref{eq:duality} and
$\stackrel{\star\star}{\Longleftrightarrow}$
is by Lemma~\ref{lem:pos-trace}.
\end{myproof}

In fact, we can say more about the embedding.
Notice that $I(\vec{n})$ is a finite dimensional
$W^*$-algebra for each $\vec{n}\in \CPM$.
Hence the embedding is restricted to
$I_{\mathrm{fd}}\colon\CPM\to(\FdWstar_\CP)^\op$,
where $\FdWstar_\CP$ denotes the category
of finite dimensional $W^*$-algebras and
normal completely positive maps.
In the same way we have an embedding
$I'_{\mathrm{fd}}\colon\Q\to(\FdWstar_\CPPU)^\op$.

\begin{mytheorem}
The embeddings
\[
I_{\mathrm{fd}}\colon\CPM\to(\FdWstar_\CP)^\op
\enspace,\qquad
I'_{\mathrm{fd}}\colon\Q\to(\FdWstar_\CPPU)^\op
\]
give equivalences of categories:
\[
\CPM\simeq(\FdWstar_\CP)^\op
\enspace,\qquad
\Q\simeq(\FdWstar_\CPPU)^\op
\enspace.
\]
\end{mytheorem}
\begin{myproof}
\cite[Theorem~I.11.2]{Takesaki2002}
implies that
$I_{\mathrm{fd}}$ and $I'_{\mathrm{fd}}$
are essentially surjective.
A full, faithful and essentially surjective functor
is a part of equivalence~\cite[Theorem~IV.4.1]{MacLane1998}.
\end{myproof}


\section{Infinite types
and classical computation}
\label{sect:infinite-type}

Because Selinger's category $\Q$ is (contravariantly)
embedded into $\Wstar_\CPPU$,
the category $\Wstar_\CPPU$ can be thought of as
an infinite dimensional extension of $\Q$.
Therefore, working in the category $\Wstar_\CPPU$
rather than $\Q$ enables us to handle
infinite types.

Recall that a type $\mathbf{bit}$
is interpreted by $\sem{\mathbf{bit}}=\C\oplus\C$.
It is easily extended to
an interpretation of
$\mathbf{trit}$ by
$\sem{\mathbf{trit}}=\C\oplus\C\oplus\C$,
and in general,
an interpretations of a type of $n$-level classical system
by $\bigoplus_{i=1}^n\C$.
Then, it is natural to interpret
a type $\mathbf{nat}$ of natural numbers by
$\sem{\mathbf{nat}}=\bigoplus_{i\in\N}\C$,
as Selinger also suggested in~\cite[\S7.3]{Selinger2004}.
Actually we have infinite direct sums
in $\Wstar_\CPPU$, but not in $\Q$.

We can also consider quantum analogue.
Interpretations $\sem{\mathbf{qbit}}=\mat_2\cong\bsp(\C^2)$,
$\sem{\mathbf{qtrit}}=\mat_3\cong\bsp(\C^3)$,
$\dotsc$, could be extended to
an interpretation
of a type of countable level quantum system
(or ``quantum natural numbers''?)
by $\bsp(\calH_\N)$,
where $\calH_\N$ is the Hilbert space of countable dimension.
This type cannot be interpreted in $\Q$.

In what follows,
we will generalise the former
observation on classical types.
\begin{mydefinition}
For a set $S$
and for a real number $p\ge 1$,
we define
\[
\ell^p(S)
\coloneqq
\Bigl\{\varphi\colon S\to \C
\Bigm|
\sum_{s\in S}\abs{\varphi(s)}^p
<\infty
\Bigr\}
\enspace, \qquad
\ell^\infty(S)
\coloneqq
\Bigl\{\varphi\colon S\to \C
\Bigm|
\sup_{s\in S}\abs{\varphi(s)}
<\infty
\Bigr\}
\enspace.
\]
It is a standard fact that
these are Banach spaces with
coordinatewise operations, and
norms $\norm{\varphi}_p=(\sum_{s\in S}\abs{\varphi(s)}^p)^{1/p}$
and $\norm{\varphi}_\infty=\sup_{s\in S}\abs{\varphi(s)}$
respectively.
\end{mydefinition}
Then, notice that
$\sem{\mathbf{bit}}=\C\oplus\C\cong\ell^\infty(2),
\sem{\mathbf{trit}}=\C\oplus\C\oplus\C\cong\ell^\infty(3),
\dotsc,\sem{\mathbf{nat}}=\bigoplus_{i\in\N}\C\cong\ell^\infty(\N)$.
We have the following general result.
\begin{myproposition}
\label{prop:countable-comm-wstar}
Let $S$ and $T$ be sets.
\begin{enumerate}
\item\label{lem-enum:ell-infi-wstar}
$\ell^\infty(S)$ is a $W^*$-algebra with the predual $\ell^1(S)$.
\item\label{lem-enum:ell-infi-starhom}
Any function $f\colon S\to T$
induces a normal unital $*$-homomorphism
$\ell^\infty(f)\colon \ell^\infty(T)\to \ell^\infty(S)$
by $\ell^\infty(f)(\varphi)=\varphi\circ f$.
\item\label{lem-enum:wtensor-and-prod}
There is a (normal unital) $*$-isomorphism:
$\ell^\infty(S)\wtensor\ell^\infty(T)\cong
\ell^\infty(S\times T)$.
\end{enumerate}
\end{myproposition}
\begin{myproof}
\ref{lem-enum:ell-infi-wstar}:
It is straightforward to see $\ell^\infty(S)$
is a $C^*$-algebra.
The duality $\ell^\infty(S)\cong\ell^1(S)^*$
is standard.\footnote{%
A really standard result is the case where $S=\N$,
but its proof is easily generalised.
Alternatively, one can
think of $\ell^\infty(S)\cong\ell^1(S)^*$
as a special case of $L^\infty(X,\Sigma,\mu)\cong L^1(X,\Sigma,\mu)^*$
for a localisable measure space $(X,\Sigma,\mu)$
(see e.g.~\cite[\S243]{Fremlin2010vol2}),
using the counting measure on $S$.}
\auxproof{%
To see $\ell^1(S)$ is the predual of $\ell^\infty(S)$,
define a map $\iota\colon\ell^\infty(S)\to\ell^1(S)^*$
by $\iota(\varphi)(\psi)=\sum_{s\in S}\varphi(s)\psi(s)$.
Then it is straightforward to check this map is an isometry.
It is also surjective as follows.
For any $\Psi\in\ell^1(S)^*$,
$\Psi\circ \delta$ is in $\ell^\infty(S)$,
where $\delta\colon S\to \ell^1(S)$ is
defined by $\delta(s)(s)=1$ and
$\delta(s)(s')=0$ ($s\ne s'$).
To prove $\iota(\Psi\circ\delta)=\Psi$,
it suffices to show $\iota(\Psi\circ\delta)(\psi)=\Psi(\psi)$
for all $\psi\colon S\to\C$ with finite support,
because they form a dense subset of $\ell^1(S)$.
For $\psi\colon S\to\C$ with finite support $F\subseteq S$,
\begin{align*}
\iota(\Psi\circ\delta)(\psi)
&= \sum_{s\in S}(\Psi\circ\delta)(s)\psi(s) \\
&= \sum_{s\in S}\Psi(\delta(s))\psi(s) \\
&= \sum_{s\in F}\Psi(\delta(s))\psi(s) \\
&= \sum_{s\in F}\Psi(\psi(s)\delta(s)) \\
&= \Psi\Bigl(\sum_{s\in F}\psi(s)\delta(s)\Bigr) \\
&= \Psi(\psi)
\enspace.
\end{align*}
}

\ref{lem-enum:ell-infi-starhom}:
It is easy to see $\ell^\infty(f)$ is a unital $*$-homomorphism.
To see it is normal,
we can give explicitly the predual map $\ell^1(f)\colon \ell^1(S)\to\ell^1(T)$
by $\ell^1(f)(\varphi)(t)=\sum_{s\in f^{-1}(t)}\varphi(s)$,
which is bounded linear, and makes the following diagram commute.
\[
\xymatrix@C=3pc{
\ell^\infty(T) \ar[d]_\cong
\ar[r]^{\ell^\infty(f)} &
\ell^\infty(S) \ar[d]^\cong \\
\ell^1(T)^*\ar[r]^{\ell^1(f)^*} & \ell^1(S)^*
}
\]

\ref{lem-enum:wtensor-and-prod}:
First we will show $\ell^1(S)\otimes\ell^1(S)\cong
\ell^1(S\times T)$, i.e.\ the preduals are isometrically isomorphic.
It is known that, when at least one of $W^*$-algebras $M,N$
is commutative, the dual spatial $C^*$-norm on $M_*\odot N_*$ coincides with
the projective (i.e.\ greatest) cross norm~\cite[\S1.22]{Sakai1998}.
Hence $\ell^1(S)\otimes\ell^1(T)$ is the projective tensor product
of $\ell^1(S)$ and $\ell^1(T)$, which possesses the universality
to represent bounded bilinear maps
$\ell^1(S)\times\ell^1(T)\to X$
(see \cite[Theorem~2.9]{Ryan2002} for detail).
Therefore, to prove $\ell^1(S)\otimes\ell^1(S)\cong
\ell^1(S\times T)$, it suffices to show the following:
there is a short bilinear map
$\eta\colon\ell^1(S)\times\ell^1(T)\to\ell^1(S\times T)$
and for each short bilinear map
$f\colon\ell^1(S)\times\ell^1(T)\to X$,
there exists a unique
short map
$\bar{f}\colon\ell^1(S\times T)\to X$
such that $f=\bar{f}\circ\eta$.\footnote{
A bilinear map $f$ is short if $\norm{f(x,y)}\le\norm{x}\norm{y}$;
and a linear map $g$ is short if $\norm{g(x)}\le\norm{x}$.}
We define $\eta(\varphi,\psi)(s,t)=\varphi(s)\psi(t)$,
then it is easy to check $\eta$ is short bilinear.
Let $f\colon\ell^1(S)\times\ell^1(T)\to X$
be a short bilinear map.
Let
\[
F(S\times T)=\{\varphi\colon S\times T\to \C\mid
\text{$\varphi$ has finite support}\}
\]
be a vector space,
which is a dense subspace of $\ell^1(S\times T)$.
Note that $\{\delta(s,t)\mid s\in S, t\in T\}$
is a basis of this space,
where $\delta\colon S\times T\to F(S\times T)$
is Kronecker's delta.
Then we define a linear map $f'\colon F(S\times T)\to X$ by
$f'(\delta(s,t))=f(\delta(s),\delta(t))$.
With respect to $\ell^1$-norm, it is short because
\begin{align*}
\norm[\Big]{f\Bigl(\sum_{i=1}^n a_i \delta(s_i,t_i)\Bigr)}
&= \norm[\Big]{\sum_{i=1}^n a_i f(\delta(s_i),\delta(t_i))} \\
&\le \sum_{i=1}^n \abs{a_i} \norm{f(\delta(s_i),\delta(t_i))} \\
&\le \sum_{i=1}^n \abs{a_i} \norm{\delta(s_i)}\norm{\delta(t_i)} \\
&= \sum_{i=1}^n \abs{a_i}
= \norm[\Big]{\sum_{i=1}^n a_i \delta(s_i,t_i)}
\enspace.
\end{align*}
Hence by density $f'$ extends to
a short map $\bar{f}\colon \ell^1(S\times T)\to X$.
From $\bar{f}(\eta(\delta(s),\delta(t)))
=\bar{f}(\delta(s,t))=f(\delta(s),\delta(t))$,
(bi)linearity and continuity, it follows that
$\bar{f}\circ \eta=f$,
and that such $\bar{f}$ is unique.
We conclude that $\ell^1(S)\otimes\ell^1(T)\cong\ell^1(S\times T)$.

By dualising, we obtain an isomorphism
\[
\ell^\infty(S)\wtensor\ell^\infty(T)\cong
(\ell^1(S)\otimes\ell^1(T))^*\cong
\ell^1(S\times T)^*\cong
\ell^\infty(S\times T)
\enspace.
\]
Finally we have to show this isomorphism is a
$*$-isomorphism.
Let $\Theta\colon \ell^\infty(S)\odot\ell^\infty(T)\to \ell^\infty(S\times T)$
be a map defined by $\Theta(\varphi\otimes\psi)(s,t)=\varphi(s)\varphi(t)$.
Then it is easy to check $\Theta$ is a $*$-homomorphism.
Now it suffices to show the following diagram commutes,
\[
\xymatrix{
\ell^\infty(S)\odot\ell^\infty(T)
\ar@{>->}[d] \ar[dr]^\Theta & \\
\ell^\infty(S)\wtensor\ell^\infty(T)
\ar[r]_-\cong&
\ell^\infty(S\times T)
}
\]
because the canonical embedding
$\ell^\infty(S)\odot\ell^\infty(T)\to
\ell^\infty(S)\wtensor\ell^\infty(T)$
is ultraweakly dense.
Checking this commutativity is straightforward.
\auxproof{(This is an old proof.)
Note that $\ell^2(S)$ is a Hilbert space, and
$\ell^\infty(S)$ has a canonical normal
unital faithful representation
$\pi\colon\ell^\infty(S)\to\bsp(\ell^2(S))$
by $\pi(\varphi)(\psi) = \varphi\psi$
(pointwise multiplication).
We can identify the spatial $W^*$-tensor product $\wtensor$
with the tensor product of von Neumann algebras.
Note the isomorphism
$\ell^2(S)\otimes\ell^2(T)\cong \ell^2(S\times T)$
of Hilbert spaces.
By the identification $\bsp(\ell^2(S)\otimes\ell^2(T))\cong
\bsp(\ell^2(S\times T))$,
we have an inclusion
$\ell^\infty(S)\odot\ell^\infty(T)\subseteq
\ell^\infty(S\times T)$.\footnote{%
Here the symbol $\odot$ denotes the algebraic tensor product.}
The (ultra)weak denseness of the inclusion
proves $\ell^\infty(S)\wtensor\ell^\infty(T)\cong
\ell^\infty(S\times T)$.}
\end{myproof}

\begin{mycorollary}
\label{cor:embed-set-wstar}
There is an embedding
$\ell^\infty\colon\Set\to(\CWstar_\MIU)^\op$,
from the category of sets and functions
to the category of commutative $W^*$-algebras
and unital $*$-homomorphisms.
Moreover, it maps binary product of sets to
the spatial $W^*$-tensor product.
\qed
\end{mycorollary}
\begin{myproof}
By Proposition~\ref{prop:countable-comm-wstar},
$\ell^\infty$ defines a functor which
maps binary product of sets to
the spatial $W^*$-tensor product.
The injectivity on objects is obvious.
We show it is faithful.
Let $f,g\colon S\to T$ be functions with
$\ell^\infty(f)=\ell^\infty(g)$.
Let $\delta\colon T\to\ell^\infty(T)$ be
Kronecker's delta.
Then,
for each $s\in S$,
\[
\delta(f(s))(g(s))
=\ell^\infty(g)(\delta(f(s)))(s)
=\ell^\infty(f)(\delta(f(s)))(s)
=\delta(f(s))(f(s))
=1
\enspace,
\]
so that $f(s)=g(s)$.
Hence $f=g$.
\end{myproof}

As a consequence, we can embed
sets $S_1,\dotsc,S_n,T$
and a function
$f\colon S_1\times\dotsb\times S_n\to T$
contravariantly
into the category $\CWstar_\MIU$
(hence into $\Wstar_\CPPU$)
as:
\[
\ell^\infty(f)\colon
\ell^\infty(T)
\longto
\ell^\infty(S_1\times\dotsb\times S_n)
\cong
\ell^\infty(S_1)
\wtensor\dotsb\wtensor
\ell^\infty(S_n)
\enspace.
\]
Therefore,
any classical data type
and function between them
interpreted in $\Set$
can also be interpreted
in $\Wstar_\CPPU$.
This enables us, for example,
to build in
a function symbol
$f\colon \mathbf{nat}^{\otimes n}\to \mathbf{nat}$
with an (classical) interpretation $\sem{f}\colon\N^n\to \N$
into a quantum programming language.

\section{Conclusions and future work}
\label{sec:concl-future}

We have given a novel denotational semantics
for a first-order functional quantum
programming language QPL
by operator algebras.
Technically, we showed that
the category $\Wstar_\CPPU$ of
$W^*$-algebras and normal completely positive
subunital maps is
a $\Dcppobot$-enriched symmetric monoidal category
with $\Dcppobot$-enriched finite products,
and therefore its opposite is an
elementary quantum flow chart category.
Our model can be considered as an infinite dimensional
extension of Selinger's original model $\Q$,
and is more flexible model
to accommodate infinite structures
and to unify classical computation
as commutative structures.

In parallel with the present work,
Rennela~\cite{Rennela2014} recently
showed that $\Wstar_\PPU$ is algebraically compact for
a certain class of (``von Neumann'') functors.
These results demonstrate that
the use of operator algebras, especially $W^*$-algebras,
for modelling quantum computation
is quite promising.
One of the most important future work is
to give a denotational semantics
by operator algebras
for a \emph{higher-order} quantum programming language,
or the quantum lambda calculus.
In an (unpublished) paper~\cite{Kornell2012},
it is showed that
the symmetric monoidal category
$((\Wstar_\MIU)^\op,\wtensor,\C)$
is \emph{closed}. This result can be helpful.

Another future work is,
extending the results in
\S\ref{sect:infinite-type}, to study
commutative $W^*$-algebras
for modelling classical computation.
It is known that
commutative $W^*$-algebras
are characterised as
$L^\infty$ spaces $L^\infty(X,\Sigma,\mu)$
of localisable measure spaces $(X,\Sigma,\mu)$
(see e.g.~\cite[\S1.18]{Sakai1998}).
Hence, by analogy with
the categorical (Gelfand)
duality between commutative $C^*$-algebras
and compact Hausdorff spaces,
it is natural to expect the categorical duality
between commutative $W^*$-algebras and
certain measure (or measurable) spaces.
In that case, Proposition~\ref{prop:countable-comm-wstar}
and Corollary~\ref{cor:embed-set-wstar}
could be obtained as a special case for
counting measures.
We further expect a similar result to
\cite{FurberJ2013}, showing that
$\Wstar_\CPPU$ can also accommodate
probabilistic computation.

\subsection*{Acknowledgements}
This paper is based on my master
thesis~\cite{Cho2014Master} at
the University of Tokyo.
I am deeply grateful to
my supervisor Ichiro Hasuo
and colleagues at the time
for valuable discussions.
I also wish to thank
my current colleagues at Radboud University
Nijmegen and the QPL reviewers
for helpful comments.


\appendix

\section{Monoidal traces on $\omcppo$-enriched cartesian categories}
\label{sec:trace-on-omcppo-enrieched-cat}

Here we will give a proof of Theorem~\ref{thm:cppo-enriched-cart-cat}.
We in fact show the dual statement,
i.e.\ Theorem~\ref{thm:cppo-enriched-cart-cat-appendix}
and Theorem~\ref{thm:cppo-enriched-cart-funct} below,
because cartesian categories are more standard
than cocartesian categories.

First, let us clarify terminology here.
A \emph{cartesian category}
is a (symmetric) monoidal category whose
monoidal structure is given by finite products.
In other words, it is just a category
with (a choice of) finite products.
A functor between cartesian categories
is \emph{cartesian} if it preserves finite products.
Note that the cartesian product functor $\times$
of an $\omcppo$-enriched cartesian category
is required to be $\omcppo$-enriched,
or equivalently, the tupling
$\tuple{\cdot,\cdot}$ be $\omega$-continuous.

The following is the first theorem we wish to prove.
\begin{mytheorem}
\label{thm:cppo-enriched-cart-cat-appendix}
Every $\omcppo$-enriched cartesian category
with left-strict composition
(i.e.\ $\bot\circ f=\bot$)
is traced.
For $f\colon A\times X\to B\times X$,
the trace $\Tr(f)\colon A\to B$ is given by
\begin{equation}
\label{eq:Tr-def}
\Tr(f)\coloneqq
\pi_1\circ
\bigvee_{n\in \N}
\Trn{n}(f)
\enspace,
\end{equation}
where $\Trn{n}(f)\colon A\to B\times X$ is defined by
\begin{align*}
\Trn{0}(f)&=\bot \\
\Trn{n+1}(f)&=f\circ \tuple[\big]{\id_A,\pi_2\circ \Trn{n}(f)}
\enspace.
\end{align*}
\end{mytheorem}
We use the well-known theorem
of Hasegawa and Hyland.
The complete proof is found
in Hasegawa's thesis~\cite[Theorem~7.1.1]{Hasegawa1999}.
\begin{mytheorem}[Hasegawa/Hyland]
\label{monoidal-trace-and-conway-op}
A cartesian category is traced if and only if
it has a Conway operator.
A trace operator $\Tr$ and a Conway operator
$\Fix$ are related bijectively as follows:
\begin{equation}
\label{eq:Tr-from-Fix}
\Tr(f) = \pi_1\circ f\circ \tuple[\big]{\id_A, \Fix(\pi_2\circ f)}
\end{equation}
for $f\colon A\times X\to B\times X$, and
\[
\Fix(g) = \Tr(\Delta_X\circ g)
\]
for $g\colon A\times X\to X$,
where $\Delta_X=\tuple{\id_X,\id_X}$ is the diagonal map.
\qed
\end{mytheorem}
Thanks to this theorem,
the problem is reduced to a little easier problem on
a Conway operator.
The following result is already showed
in~\cite[Lemma~A.1 in Appendix]{HoshinoMH2014LICS}.
\begin{mytheorem}
\label{cpo-enriched-cat-has-conway}
Every $\omcppo$-enriched cartesian category
with left-strict composition
(i.e.\ $\bot\circ f=\bot$)
has a Conway operator $\Fix$.
For $g\colon A\times X\to X$,
$\Fix(g)\colon A\to X$ is given by
\begin{equation}
\label{eq:Fix-def}
\Fix(g)\coloneqq
\bigvee_{n\in \N}
\Fixn{n}(g)
\enspace,
\end{equation}
where $\Fixn{n}(g)\colon A\to X$ is defined by
\begin{align*}
\Fixn{0}(g)&=\bot \\
\Fixn{n+1}(g)&=g\circ \tuple[\big]{\id_A,\Fixn{n}(g)}
\enspace.
\end{align*}
\qed
\end{mytheorem}
Now we prove Theorem~\ref{thm:cppo-enriched-cart-cat-appendix} as follows.
\begin{myproof}[Proof of Theorem~\ref{thm:cppo-enriched-cart-cat-appendix}]
Let $\catC$ be an $\omcppo$-enriched cartesian category
with left-strict composition.
Then,
by Theorem~\ref{monoidal-trace-and-conway-op}
and
Theorem~\ref{cpo-enriched-cat-has-conway},
$\catC$ is traced.
We still need to check the equation \eqref{eq:Tr-def}.
For $f\colon A\times X\to B\times X$,
\begin{align*}
\Tr(f) &= \pi_1\circ f\circ \tuple[\big]{\id_A, \Fix(\pi_2\circ f)}
&&\text{by \eqref{eq:Tr-from-Fix}}
\\
&=
\pi_1\circ f\circ \tuple[\Big]{\id_A,
\bigvee_{n\in \N}
\Fixn{n}(\pi_2\circ f)}
&&\text{by \eqref{eq:Fix-def}}
\\
&=
\pi_1\circ
\bigvee_{n\in \N}
\Bigl(
f\circ \tuple[\big]{\id_A,
\Fixn{n}(\pi_2\circ f)}
\Bigr)
\enspace.
\end{align*}
It is easy to see, by induction on $n$,
\[
\Trn{n}(f) \le
f\circ \tuple[\big]{\id_A,
\Fixn{n}(\pi_2\circ f)}
\le
\Trn{n+1}(f)
\]
for all $n\in\N$.
It follows that
\[
\bigvee_{n\in \N}
\Bigl(
f\circ \tuple[\big]{\id_A,
\Fixn{n}(\pi_2\circ f)}
\Bigr)
=
\bigvee_{n\in \N}
\Trn{n}(f)
\enspace.
\]
Hence we have
\[
\Tr(f) =
\pi_1\circ
\bigvee_{n\in \N}
\Trn{n}(f)
\enspace.
\]
\end{myproof}

Next, we will show the theorem on cartesian functors.
\begin{mytheorem}
\label{thm:cppo-enriched-cart-funct}
Let $\catC$ and $\catD$
be $\omcppo$-enriched cartesian categories,
which are traced by
Theorem~\ref{thm:cppo-enriched-cart-cat-appendix}.
Then, every
$\omcppo$-enriched cartesian functor
between $\catC$ and $\catD$
satisfying $F\bot=\bot$
is traced.
\end{mytheorem}
\begin{myproof}
Specifically, we need to show the equation
\[
F\Tr(f) = \Tr(\phi_B\circ Ff\circ \phi_A^{-1})
\]
for $f\colon A\times X\to B\times X$ in $\catC$,
where
\begin{gather*}
\phi_A\coloneqq\tuple{F\pi_1,F\pi_2}\colon F(A\times X)\stackrel{\cong}{\longto} FA\times FX \\
\phi_B\coloneqq\tuple{F\pi_1,F\pi_2}\colon F(B\times X)\stackrel{\cong}{\longto} FB\times FX
\end{gather*}
are the canonical isomorphisms.
Note first that we have the following commutative diagrams.
\[
\xymatrix{
& \ar[dl]_{F\pi_1} F(A\times X)\ar[d]^{\phi_A}_{\cong} \ar[dr]^{F\pi_2} & \\
FA & \ar[l]^-{\pi_1} FA\times FX \ar[r]_-{\pi_2} & FX
}
\qquad
\xymatrix{
FW \ar[d]_{F\tuple{k,l}} \ar[dr]^{\tuple{Fk,Fl}} & \\
F(A\times X)\ar[r]_{\phi_A}^{\cong} & FA\times FX
}
\]
In the right diagram,
$k\colon W\to A$ and $l\colon W\to X$ are arbitrary arrows in $\catC$.
Similar diagrams commute for $\phi_B$.
Then,
\begin{align*}
F\Tr(f)
&=
F\Bigl(\pi_1\circ
\bigvee_{n\in \N}
\Trn{n}(f)\Bigr)
\\
&=
F\pi_1\circ
\bigvee_{n\in \N}
F\Trn{n}(f)
\\
&=
\pi_1\circ \phi_B\circ
\bigvee_{n\in \N}
F\Trn{n}(f)
\\
&=
\pi_1\circ
\bigvee_{n\in \N}
\Bigl(\phi_B\circ
F\Trn{n}(f)\Bigr)
\enspace,
\end{align*}
while
\[
\Tr(\phi_B\circ Ff\circ \phi_A^{-1})
= \pi_1\circ
\bigvee_{n\in \N}
\Trn{n}(\phi_B\circ Ff\circ \phi_A^{-1})
\enspace.
\]
Hence it suffices to show
\[
\bigvee_{n\in \N}
\Bigl(\phi_B\circ
F\Trn{n}(f)\Bigr)
=
\bigvee_{n\in \N}
\Trn{n}(\phi_B\circ Ff\circ \phi_A^{-1})
\enspace.
\]
This follows from
\[
\Trn{n}(\phi_B\circ Ff\circ \phi_A^{-1})
\le
\phi_B\circ
F\Trn{n}(f)
\le
\Trn{n+1}(\phi_B\circ Ff\circ \phi_A^{-1})
\]
for all $n\in\N$.
We will prove it by induction on $n$.
\begin{enumerate}[label=(\roman*)]
\item
  Base case ($n=0$):
  \[
  \Trn{0}(\phi_B\circ Ff\circ \phi_A^{-1})
  =\bot \le
  \phi_B\circ F\Trn{0}(f)
  \]
  shows the first inequality,
  and
  \begin{align*}
  &\Trn{1}(\phi_B\circ Ff\circ \phi_A^{-1})
  \\
  &=
  \phi_B\circ Ff\circ \phi_A^{-1}\circ
  \tuple[\big]{\id_{FA},
  \pi_2\circ
  \Trn{0}(\phi_B\circ Ff\circ \phi_A^{-1})}
  \\
  &\ge
  \phi_B\circ \bot
  \\
  &=
  \phi_B\circ F\bot
  &&\text{by strictness of $F$}
  \\
  &=
  \phi_B\circ F\Trn{0}(f)
  \end{align*}
  shows the second inequality.
\item
  Induction step:
  \begin{align*}
  &\phi_B\circ
  F\Trn{n+1}(f)
  \\
  &=
  \phi_B\circ
  F\bigl(f\circ
  \tuple[\big]{\id_A,\pi_2\circ\Trn{n}(f)}
  \\
  &=
  \phi_B\circ
  Ff\circ\phi_A^{-1}\circ
  \tuple[\big]{F\id_A,F\bigl(\pi_2\circ\Trn{n}(f)\bigr)}
  \\
  &=
  \phi_B\circ
  Ff\circ\phi_A^{-1}\circ
  \tuple[\big]{\id_{FA},\pi_2\circ\phi_B\circ\Trn{n}(f)}
  \\
  &\ge
  \phi_B\circ
  Ff\circ\phi_A^{-1}\circ
  \tuple[\big]{\id_{FA},\pi_2\circ
  \Trn{n}(\phi_B\circ Ff\circ \phi_A^{-1})}
  &&\text{by I.H.}
  \\
  &=
  \Trn{n+1}(\phi_B\circ Ff\circ \phi_A^{-1})
  \end{align*}
  shows the first inequality, and similarly
  \begin{align*}
  &\phi_B\circ
  F\Trn{n+1}(f)
  \\
  &=
  \phi_B\circ
  Ff\circ\phi_A^{-1}\circ
  \tuple[\big]{\id_{FA},\pi_2\circ\phi_B\circ\Trn{n}(f)}
  \\
  &\le
  \phi_B\circ
  Ff\circ\phi_A^{-1}\circ
  \tuple[\big]{\id_{FA},\pi_2\circ
  \Trn{n+1}(\phi_B\circ Ff\circ \phi_A^{-1})}
  &&\text{by I.H.}
  \\
  &=
  \Trn{n+2}(\phi_B\circ Ff\circ \phi_A^{-1})
  \end{align*}
  shows the second inequality.
\end{enumerate}
Therefore we have
\[
F\Tr(f) = \Tr(\phi_B\circ Ff\circ \phi_A^{-1})
\enspace.
\]
\end{myproof}

\section{Distribution of tensor products over direct sums}
\label{sec:dist-tensorprod-directsum}

First we recap tensor products of $C^*$- and $W^*$-algebras.
Let $A$ and $B$ be $C^*$-algebras.
We denote by $A\odot B$ the algebraic tensor product of
$A$ and $B$. Then it is easy to see $A\odot B$ is a $*$-algebra.
A $C^*$-norm on $A\odot B$ is not necessarily unique,
but there exists a least $C^*$-norm on $A\odot B$, which is called
the \emph{spatial} (or \emph{minimal}, or \emph{injective})
\emph{$C^*$-norm}.
The \emph{spatial} (or \emph{minimal}, or \emph{injective})
\emph{$C^*$-tensor product} $A\otimes B$ of $A$ and $B$
is defined to be the completion of
$A\odot B$ under the spatial $C^*$-norm.
Later we will use the next result,
which also explains the term ``spatial''.
\begin{mytheorem}[{\cite[Theorem~IV.4.9.(iii)]{Takesaki2002}}]
Let $A,B$ be $C^*$-algebras,
and $(\calH_A,\pi_A),(\calH_B,\pi_B)$
faithful representations of $A,B$ respectively.
Then the mapping $a\otimes b\mapsto\pi_A(a)\otimes\pi_B(b)$
extends to a faithful representation
$\pi_A\otimes\pi_B\colon A\otimes B\to \bsp(\calH_A\otimes\calH_B)$
of $A\otimes B$ on $\calH_A\otimes\calH_B$.
\qed
\end{mytheorem}

We also use a similar (but much easier to prove)
result on direct sums.
\begin{mytheorem}
Let $A,B$ be $C^*$-algebras,
and $(\calH_A,\pi_A),(\calH_B,\pi_B)$
faithful representations of $A,B$ respectively.
Then the mapping $(a, b)\mapsto\pi_A(a)\oplus\pi_B(b)$
gives a faithful representation
$\pi_A\oplus\pi_B\colon A\oplus B\to \bsp(\calH_A\oplus\calH_B)$
of $A\oplus B$ on $\calH_A\oplus\calH_B$.
\qed
\end{mytheorem}

Let $M$ and $N$ be $W^*$-algebras.
Consider the algebraic tensor product
$M_*\odot N_*$ of their preduals $M_*$ and $N_*$.
Because the spatial $C^*$-norm is a cross norm,
we can equip
$M_*\odot N_*$ with the dual norm
of the spatial $C^*$-norm
via the following embedding:
\[
M_*\odot N_*\longhookrightarrow
M^*\odot N^*\longhookrightarrow
(M\otimes N)^*
\enspace.
\]
Let $M_*\otimes N_*$ denote the completion
of $M_*\odot N_*$ under this norm.
Then, it turns out that
the dual $(M_*\otimes N_*)^*$ is a $C^*$-algebra,
hence a $W^*$-algebra with the predual $M_*\otimes N_*$.
The $W^*$-algebra $(M_*\otimes N_*)^*$
is denoted by $M\wtensor N$,
and called the \emph{spatial $W^*$-tensor product}
of $M$ and $N$.
The following fact is useful:
the spatial $C^*$-tensor product
$M\otimes N$ is ultraweakly densely embedded into
$M\wtensor N$ (and therefore $M\odot N$ also forms a dense subspace of
$M\wtensor N$).
With respect to this canonical embeddings,
it is straightforward to see that
the $C^*$-tensor product of maps $f\otimes g$
(if defined) extends to
the $W^*$-tensor product $f\wtensor g=(f_*\otimes g_*)^*$.
For further details of tensor products
of $C^*$- and $W^*$-algebras,
see e.g.~\cite[\S1.22]{Sakai1998},
\cite[Chapter~IV]{Takesaki2002}.

For $p$ with $1\le p\le\infty$,
we will denote by $\oplus^p$
the $\ell^p$-direct sum of normed spaces.
Note that the direct sum
of $C^*$- and $W^*$-algebras
is the $\ell^\infty$-direct sum.
For $W^*$-algebras $(M_i)_i$,
the predual of $\bigoplus_i M_i$
is given by $\bigoplus^1_i M_{i*}$,
i.e.\ $(\bigoplus^1_i M_{i*})^*\cong\bigoplus_i M_i$
(\cite[Definition~1.1.5]{Sakai1998}).
For the finite direct sum,
we have the other result.
\begin{mylemma}
\label{lem:dual-directsum}
Let $X,Y$ be normed spaces.
Then we have an isometric isomorphism:
\[
(X\oplus^\infty Y)^*
\cong X^*\oplus^1 Y^*
\]
\qed
\end{mylemma}

\noindent
We will also use the following elementary result.
\begin{mylemma}
\label{lem:direct-sum-dense-isom}
Let $f\colon X\to X'$ and
$g\colon Y\to Y'$ be isometries with dense images
between normed spaces.
For any $p$ ($1\le p \le\infty$),
$f\oplus^p g\colon X\oplus^p Y\to X'\oplus^p Y'$
is an isometry with dense image, too.
\qed
\end{mylemma}
\auxproof{
\begin{myproof}
$f\oplus g$ is an isometry because
\[
\norm{(f\oplus g)(x,y)}
=\norm{(f(x),g(y))}
=(\norm{f(x)}^p+\norm{g(y)}^p)^{1/p}
=(\norm{x}^p+\norm{y}^p)^{1/p}
=\norm{(x,y)}
\enspace.
\]
Take arbitrary $(x',y')\in X'\oplus Y'$.
By assumption there exist sequences $(x_n)_n$
in $X$ and $(y_n)_n$ in $Y$ such that
$(f(x_n))_n$ and $(g(y_n))_n$ are convergent
to $x'$ in $X'$ and $y'$ in $Y'$, respectively.
Then a sequence $((f\oplus g)(x_n,y_n))_n=((f(x_n),g(y_n)))_n$ converges
to $(x',y')$ in $X'\oplus Y'$ because
\[
\norm{(x',y')-(f(x_n),g(y_n))}
=\norm{(x'-f(x_n),y'-g(y_n))}
=(\norm{x'-f(x_n)}^p+\norm{y'-g(y_n)}^p)^{1/p}
\to 0
\]
\end{myproof}
}%

Now, we prove the distribution of
$C^*$- and $W^*$-tensor products
over direct sums.
\begin{mytheorem}
\label{thm:cstar-tensor-distrib}
Let $A,B,C$ be $C^*$-algebras.
Then the canonical maps
\begin{gather*}
\bang\colon A\otimes 0\longto 0
\\
\tuple{\id\otimes\pi_1,\id\otimes\pi_2}\colon
A\otimes (B\oplus C)\longto
(A\otimes B)\oplus(A\otimes C)
\end{gather*}
are $*$-isomorphisms.
\end{mytheorem}
\begin{myproof}
Note that $A\odot 0\cong 0$
and the only possible norm is the trivial one.
Therefore $A\otimes 0\cong A\odot 0\cong 0$
and the first one is proved.

We will show the latter one.
Let
\begin{align*}
\iota_1&\colon A\odot (B\oplus C)\longhookrightarrow A\otimes (B\oplus C) \\
\iota_2&\colon A\odot B\longhookrightarrow A\otimes B \\
\iota_3&\colon A\odot C\longhookrightarrow A\otimes C
\end{align*}
be the dense embeddings.
Note that the algebraic tensor product distributes over
finite direct sum, and that the following diagram commutes.
\[
\xymatrix@C=4pc{
A\otimes (B\oplus C) \ar[r]^-{\tuple{\id\otimes\pi_1,\id\otimes\pi_2}} &
(A\otimes B)\oplus (A\otimes C) \\
A\odot (B\oplus C)
\ar[r]_-\cong^-{\tuple{\id\odot\pi_1,\id\odot\pi_2}}
\ar[u]^{\iota_1} &
(A\odot B)\oplus (A\odot C)
\ar[u]_{\iota_2\oplus\iota_3}
}
\]
By the universality of completion, it suffices to show
$(\iota_B\oplus\iota_C)\circ\tuple{\id\odot\pi_1,\id\odot\pi_2}$
is an isometry with dense image
w.r.t.\ the spatial $C^*$-norm on $A\odot(B\oplus C)$.
By Lemma~\ref{lem:direct-sum-dense-isom},
$\iota_B\oplus\iota_C$ is an isometry with dense image
w.r.t.\ the spatial $C^*$-norms on $A\odot B$ and
$A\odot C$.
Hence, we only need to prove
$\tuple{\id\odot\pi_1,\id\odot\pi_2}$ is an isometry.
Let $(\calH_A, \pi_A),(\calH_B, \pi_B),(\calH_C, \pi_C)$
be faithful representations of $A,B,C$ respectively.
Then we have a faithful representation
$(\calH_B\oplus\calH_C,\pi_B\oplus\pi_C)$ of $B\oplus C$,
and a faithful representation
$(\calH_A\otimes(\calH_B\oplus\calH_C),\pi_A\otimes(\pi_B\oplus\pi_C))$
of $A\otimes(B\oplus C)$.
On the other hand, $(\calH_A\otimes\calH_B,\pi_A\otimes\pi_B)$
and $(\calH_A\otimes\calH_C,\pi_A\otimes\pi_C)$ is faithful
representations of $A\otimes B$ and $A\otimes C$ respectively,
and hence $((\calH_A\otimes\calH_B)\oplus(\calH_A\otimes\calH_C),
(\pi_A\otimes\pi_B)\oplus(\pi_A\otimes\pi_C))$ is a faithful representation
of $(A\otimes B)\oplus(A\otimes C)$.
Then, to prove
$\tuple{\id\odot\pi_1,\id\odot\pi_2}$ is an isometry,
it suffices to show the following diagram commutes
(because the other maps are isometries).
\[
\xymatrix@C=3pc{
A\odot (B\oplus C) \ar[d]^{\tuple{\id\odot\pi_1,\id\odot\pi_2}}
\ar[r]^-{\iota_1} &
A\otimes (B\oplus C)
\ar[rr]^-{\pi_A\otimes(\pi_B\oplus\pi_C)} &&
\bsp(\calH_A\otimes(\calH_B\oplus\calH_C))
\ar[d]_\cong^{\Phi} \\
(A\odot B)\oplus (A\odot C) \ar[r]^{\iota_2\oplus\iota_3} &
(A\otimes B)\oplus (A\otimes C)
\ar[rr]^-{(\pi_A\otimes\pi_B)\oplus(\pi_A\otimes\pi_C)} &&
\bsp((\calH_A\otimes\calH_B)\oplus(\calH_A\otimes\calH_C))
}
\]
Here $\Phi$ is a $*$-isomorphism induced by
an isomorphism $\phi\colon\calH_A\otimes(\calH_B\oplus\calH_C)
\to(\calH_A\otimes\calH_B)\oplus(\calH_A\otimes\calH_C)$
of Hilbert spaces. For $a\in A,b\in B,c\in C$,
\begin{align*}
&(\Phi\circ(\pi_A\otimes(\pi_B\oplus\pi_C))\circ\iota_1)(a\otimes (b,c)) \\
&=\Phi(\pi_A(a)\otimes(\pi_B(b)\oplus\pi_C(c))) \\
&=\phi\circ
(\pi_A(a)\otimes(\pi_B(b)\oplus\pi_C(c)))
\circ\phi^{-1} \\
&\stackrel{\star}{=}(\pi_A(a)\otimes\pi_B(b))\oplus(\pi_A(a)\otimes\pi_C(c)) \\
&= ((\pi_A\otimes\pi_B)\oplus(\pi_A\otimes\pi_C))(a\otimes b,a\otimes c) \\
&= (((\pi_A\otimes\pi_B)\oplus(\pi_A\otimes\pi_C))
\circ(\iota_2\oplus\iota_3)\circ
\tuple{\id\odot\pi_1,\id\odot\pi_2})(a\otimes(b,c))
\enspace,
\end{align*}
where the marked equality $\stackrel{\star}{=}$
is showed by checking the commutativity of the following diagram.
\[
\xymatrix{
\calH_A\otimes(\calH_B\oplus\calH_C) \ar[r]_-\cong^-{\phi}
\ar[d]_{\pi_A(a)\otimes(\pi_B(b)\oplus\pi_C(c))} &
(\calH_A\otimes\calH_B)\oplus(\calH_A\otimes\calH_C)
\ar[d]^{(\pi_A(a)\otimes\pi_B(b))\oplus(\pi_A(a)\otimes\pi_C(c))} \\
\calH_A\otimes(\calH_B\oplus\calH_C) \ar[r]_-\cong^-{\phi} &
(\calH_A\otimes\calH_B)\oplus(\calH_A\otimes\calH_C)
}
\]
\end{myproof}

\begin{mytheorem}
Let $M,N,L$ be $W^*$-algebras.
Then the canonical maps
\begin{gather*}
\bang\colon M\wtensor 0\longto 0
\\
\tuple{\id\wtensor\pi_1,\id\wtensor\pi_2}\colon
M\wtensor (N\oplus L)\longto
(M\wtensor N)\oplus(M\wtensor L)
\end{gather*}
are $*$-isomorphisms.
\end{mytheorem}
\begin{myproof}
Note that $0_*\cong 0$.
Hence $M_*\otimes 0_* \cong M_*\odot 0_*\cong 0$
and $M\wtensor 0 = (M_*\otimes 0_*)^* \cong 0$,
which proves the first one.

We will show the latter one.
Note that their preduals are
explicitly given:
\begin{align*}
(M\wtensor (N\oplus L))_*
&\cong M_*\otimes (N\oplus L)_*
\cong M_*\otimes (N_*\oplus^1 L_*) \\
((M\wtensor N)\oplus(M\wtensor L))_*
&\cong (M\wtensor N)_*\oplus^1(M\wtensor L)_*
\cong (M_*\otimes N_*)\oplus^1(M_*\otimes L_*)
\end{align*}
So, we will first show $M_*\otimes (N_*\oplus^1 L_*)
\cong (M_*\otimes N_*)\oplus^1(M_*\otimes L_*)$.
For this, by a similar argument to
the proof of Theorem~\ref{thm:cstar-tensor-distrib},
it suffices to show
$\tuple{\id\odot\pi_1,\id\odot\pi_2}\colon
M_*\odot(N_*\oplus^1 L_*)\to
(M_*\odot N_*)\oplus^1(M_*\odot L_*)$
is an isometry.
This follows from the commutativity of the following diagram,
using Lemma~\ref{lem:dual-directsum}.
\[
\xymatrix{
M_*\odot(N_*\oplus^1 L_*)
\ar[d]^\cong_{\tuple{\id\odot\pi_1,\id\odot\pi_2}}
\ar[r]^-\cong &
M_*\odot(N\oplus L)_* \ar[r] &
(M\otimes(N\oplus L))^*
\\
(M_*\odot N_*)\oplus^1(M_*\odot L_*)
\ar[r] &
(M\otimes N)^*\oplus^1(M\otimes L)^*
\ar[r]^{\cong} &
((M\otimes N)\oplus (M\otimes L))^*
\ar[u]^\cong_{\tuple{\id\otimes\pi_1,\id\otimes\pi_2}^*}
}
\]
Hence we have an isometric isomorphism
$\tuple{\id\otimes\pi_1,\id\otimes\pi_2}
\colon M_*\otimes (N_*\oplus^1 L_*)
\to (M_*\otimes N_*)\oplus^1(M_*\otimes L_*)$.
Now, we can see the canonical map
$\tuple{\id\wtensor\pi_1,\id\wtensor\pi_2}\colon
M\wtensor (N\oplus L)\to
(M\wtensor N)\oplus(M\wtensor L)$,
i.e.\ $\tuple{(\id_*\otimes\pi_{1*})^*,(\id_*\wtensor\pi_{2*})^*}\colon
(M_*\otimes (N\oplus L)_*)^*\to
(M_*\otimes N_*)^*\oplus(M_*\otimes L_*)^*$,
is an isomorphism because we can check the following
diagram commutes.
\[
\xymatrix@C=7pc{
(M_*\otimes (N\oplus L)_*)^*
\ar[r]^-{\tuple{(\id_*\otimes\pi_{1*})^*,(\id_*\wtensor\pi_{2*})^*}}
\ar[d]_\cong &
(M_*\otimes N_*)^*\oplus(M_*\otimes L_*)^* \ar[d]^\cong \\
(M_*\otimes (N_*\oplus^1 L_*))^* &
\ar[l]_-\cong^-{\tuple{\id\otimes\pi_1,\id\otimes\pi_2}^*}
((M_*\otimes N_*)\oplus^1(M_*\otimes L_*))^*
}
\]
\end{myproof}

\section{Other omitted proofs}
\label{sec-apx:other-omitted-proofs}

\subsection{Miscellaneous results on $C^*$- and $W^*$-algebras}

Here we show miscellaneous results on $C^*$- and $W^*$-algebras,
which is used in the main text or
later in~\S\ref{subsec:proof-for-enrichment}.
Note again that, in this paper, $C^*$-algebras are always
assumed to be unital.

For an element $x\in A$ of a $C^*$-algebra,
we write $\spec(x) \coloneqq
\{\lambda\in \C \mid
x-\lambda 1\;\;\text{is not invertible}\}$
for the spectrum of $x$.
Here are several facts on spectrum in $C^*$-algebras.
\begin{myproposition}[{\cite[Proposition~I.4.2]{Takesaki2002}}]
\label{prop:norm-equal-sprad}
Let $A$ be a $C^*$-algebra and
$x\in A$ a normal element.
Then the norm of $x$ coincides with the spectral
radius of $x$, that is,
$\norm{x}=\sup_{\lambda\in\spec(x)} \abs{\lambda}$.
\qed
\end{myproposition}
\begin{myproposition}[{\cite[Proposition~I.4.3 and Theorem~I.6.1]{Takesaki2002}}]
\label{prop:spec-pos}
Let $A$ be a $C^*$-algebra and
$x\in A$ a self-adjoint element.
Then $\spec(x)\subseteq\R$; and
$\spec(x)\subseteq\R^+$ if and only if $x$ is positive.
\qed
\end{myproposition}

\begin{myproposition}
\label{prop:norm-vs-order}
Let $A$ be a $C^*$-algebra
and $x\in A$ a self-adjoint element.
For any $M\in\R^+$, we have
\[
\norm{x}\le M
\iff
-M 1\le x\le M 1
\enspace.
\]
\end{myproposition}
\begin{myproof}
Note first that for any $\alpha,\beta\in\C$ ($\alpha\ne 0$)
we have $\spec(\alpha x+\beta 1) = \alpha\cdot\spec(x)+\beta$.
Then
\begin{align*}
\norm{x}\le M
&\iff
\sup_{\lambda\in\spec(x)}\abs{\lambda}\le M
&& \text{by Proposition~\ref{prop:norm-equal-sprad}}
\\
&\iff
\forall \lambda\in \spec(x)\ldotp
-M\le \lambda \le M
&&\text{by Proposition~\ref{prop:spec-pos}}
\\
&\iff
\forall \lambda\in \spec(x)\ldotp
M-\lambda\ge 0 \text{ and }
\lambda + M\ge 0
\\
&\iff
\spec(M1-x)\subseteq\R^+
\text{ and }
\spec(x+M1)\subseteq\R^+
\\
&\iff
M1-x\ge 0
\text{ and }
x+M1\ge 0
&&\text{by Proposition~\ref{prop:spec-pos}}
\\
&\iff
-M 1\le x\le M 1
\end{align*}
\end{myproof}

\noindent
As a special case we have
$-\norm{x}1\le x \le \norm{x} 1$
for any self-adjoint element $x$.
Now the next proposition is
an easy consequence.
\begin{myproposition}
\label{prop:norm-bdd-vs-order-bdd}
Let $A$ be a $C^*$-algebra and
$S\subseteq \sa{A}$ be a set of
self-adjoint elements of $A$.
Then $S$ is norm-bounded if and only if
$S$ is order-theoretically bounded in $\sa{A}$.
\qed
\end{myproposition}

We will show some results on
the spatial $W^*$-tensor products.
\begin{mylemma}
\label{lem:fin-dim-tensor-wstar}
Let $A$ be a finite dimensional $W^*$-algebra
(note that $A_*=A^*$),
and let $M$ be a $W^*$-algebra.
Then, the algebraic tensor product
$A^*\odot M_*$ is already complete under the dual
spatial $C^*$-norm.
Moreover,
the canonical embedding
\[
A\odot M \longhookrightarrow
(A^*\odot M_*)^*
\]
is surjective, so that
$A\odot M \cong (A^*\odot M_*)^*$.
Therefore, $A\odot M$ is a $W^*$-algebra
with the predual $A^*\odot M_*$.
\end{mylemma}
\begin{myproof}
Fix a normalised basis $\{a_1,\dotsc,a_n\}$ of $A$.
We denote its dual basis by $\{\hat{a}_1,\dotsc,\hat{a}_n\}$,
which is a basis of $A^*$.
Note that every element $\chi\in A^*\odot M_*$
is uniquely written as $\chi=\sum_{i=1}^n \hat{a_i}\otimes \varphi_i$.
For arbitrary $x\in M$ with $\norm{x}\le 1$,
we have $\norm{a_i\otimes x}=\norm{a_i}\norm{x}=\norm{x}\le 1$
because any $C^*$-norm is a cross-norm.
Hence
\begin{align*}
\abs{\varphi_i(x)}
=\abs[\Big]{\Bigl(\sum_{i=1}^n \hat{a_i}\otimes \varphi_i\Bigr)(a_i\otimes x)}
=\abs{\chi(a_i\otimes x)}
\le\sup\{\abs{\chi(z)}\mid
z\in A\otimes M, \norm{z}\le 1\} = \norm{\chi}
\enspace,
\end{align*}
so that $\norm{\varphi_i}=\sup\{\abs{\varphi_i(x)}\mid
x\in M, \norm{x}\le 1\}\le \norm{\chi}$ for each $i$.
Now, assume that $(\chi_j)_j=(\sum_{i=1}^n \hat{a_i}\otimes \varphi_{ij})$
is a Cauchy sequence in $A^*\odot M_*$.
Because
\[
\norm{\varphi_{ik}-\varphi_{ij}}
\le \norm[\Big]{\sum_{i=1}^n \hat{a_i}\otimes (\varphi_{ik}-\varphi_{ij})}
= \norm{\chi_k-\chi_j}
\enspace,
\]
$(\varphi_{ij})_j$ is a Cauchy sequence for each $i$.
Let $\varphi_i=\lim_{j\to\infty}\varphi_{ij}$
and $\chi=\sum_{i=1}^n \hat{a}_i\otimes\varphi_i$.
Then
\[
\norm{\chi-\chi_j}
= \norm[\Big]{\sum_{i=1}^n \hat{a_i}\otimes (\varphi_i-\varphi_{ij})}
\le \sum_{i=1}^n \norm{\hat{a_i}}\norm{\varphi_i-\varphi_{ij}}
\to 0
\quad\text{when $j\to\infty$}\enspace.
\]
Hence $A^*\odot M_*$ is complete.

Let $\theta\colon A\odot M\to (A^*\odot M_*)^*$ be the canonical embedding.
Take arbitrary $\Phi\in(A^*\odot M_*)^*$.
For each $i$,
define $\Phi_i\colon M_*\to\C$ by
$\Phi_i(\varphi)=\Phi(\hat{a}_i\otimes\varphi)$.
Clearly $\Phi_i$ is linear, and bounded because
\[
\norm{\Phi_i(\varphi)}=\norm{\Phi(\hat{a}\otimes\varphi)}\le
\norm{\Phi}\norm{\hat{a}\otimes\varphi}
=\norm{\Phi}\norm{\hat{a}}\norm{\varphi}
\enspace.
\]
Hence $\Phi_i\in(M_*)^*$.
Then we have $\theta(\sum_{i=1}^n a_i\otimes \iota^{-1}(\Phi_i)) = \Phi$,
where $\iota\colon M\to (M_*)^*$ is the canonical isomorphism,
because
\begin{align*}
\theta\Bigl(\sum_{i=1}^n a_i\otimes \iota^{-1}(\Phi_i)\Bigr)(\hat{a}_j\otimes\varphi)
= \sum_{i=1}^n \hat{a}_j(a_i) \varphi(\iota^{-1}(\Phi_i))
= \varphi(\iota^{-1}(\Phi_j))
= \Phi_j(\varphi) = \Phi(\hat{a}_j\otimes\varphi)
\enspace.
\end{align*}
\end{myproof}

\auxproof{
\begin{mylemma}
Let $f\colon A\to B$ be a linear (automatically normal) map between
finite dimensional $W^*$-algebras,
and $g\colon M\to N$ a normal map between
$W^*$-algebras.
Then $f\odot g\colon A\odot M\to B\odot N$
is normal.
\end{mylemma}
\begin{myproof}
It suffices to show $f^*\odot g_*\colon A^*\odot M_*\to B^*\odot N_*$
is bounded.
Fix a normalised basis $\{a_1,\dotsc,a_n\}$ of $A$
and denote its dual basis by $\{\hat{a}_1,\dotsc,\hat{a}_n\}$.
As is showed in the proof of Lemma~\ref{lem:fin-dim-tensor-wstar},
we have $\norm{\varphi_i}\le\norm{\sum_{i=1}^n \hat{a}_i \otimes \varphi_i}$
for each $i$.
Then
\begin{align*}
\norm[\Big]{(f^*\odot g_*)\Bigl(\sum_{i=1}^n \hat{a}_i \otimes \varphi_i\Bigr)}
&= \norm[\Big]{\sum_{i=1}^n f^*(\hat{a}_i)\otimes g_*(\varphi_i)} \\
&\le \sum_{i=1}^n \norm{f^*(\hat{a}_i)}\norm{g_*(\varphi_i)} \\
&\le \sum_{i=1}^n \norm{f^*}\norm{\hat{a}_i}\norm{g_*}\norm{\varphi_i} \\
&\le \norm{f^*}\Bigl(\sum_{i=1}^n\norm{\hat{a}_i}\Bigr)\norm{g_*}
\norm[\Big]{\sum_{i=1}^n \hat{a}_i \otimes \varphi_i}
\enspace.
\end{align*}
\end{myproof}
} 

For a $*$-algebra $A$
and $n\in\N$,
let $\mat_n(A)$ denote
the $*$-algebras of $n\times n$
matrices with entries from $A$.
It is known that if $A$ is a $C^*$-algebra,
then $\mat_n(A)$ is a $C^*$-algebra.
Because of the obvious $*$-isomorphism
$\mat_n(A)\cong\mat_n(\C)\odot A$,
the following is an immediate consequence of the above.
\begin{mycorollary}
\label{cor:mat-wstar-is-wstar}
If $M$ is a $W^*$-algebra, then
$\mat_n(M)$ is a $W^*$-algebra.
\auxproof{
If $f\colon M\to N$ is a normal map between $W^*$-algebras,
then $\mat_n(f)\colon \mat_n(M)\to \mat_n(N)$ is normal.}
\qed
\end{mycorollary}

\begin{mylemma}
In the setting of Lemma~\ref{lem:fin-dim-tensor-wstar},
fix a basis $\{a_1,\dotsc,a_n\}$ of $A$.
Let $(z_j)_j=(\sum_{i=1}^n a_i\otimes x_{ij})_j$ be a net in $A\odot M$,
and let $z=\sum_{i=1}^n a_i\otimes x_i\in A\odot M$.
Then, $z_j\to z$ ultraweakly in $A\odot M$ if and only if
$x_{ij}\to x_i$ ultraweakly in $M$ for all $i\in\{1,\dotsc,n\}$.
\end{mylemma}
\begin{myproof}
Assume that $z_j\to z$ ultraweakly in $A\odot M$.
It means for all $\chi\in A^*\odot M_*$
one has $\chi(z_j)\to \chi(z)$.
Then for all $i$ and for all $\varphi\in M_*$,
\[
\varphi(x_{ij}) =
(\hat{a}_i\otimes\varphi)\Bigl(\sum_{i=1}^n a_i\otimes x_{ij}\Bigr)\to
(\hat{a}_i\otimes\varphi)\Bigl(\sum_{i=1}^n a_i\otimes x_i\Bigr)
= \varphi(x_i)
\enspace,
\]
that is, $x_{ij}\to x_i$ ultraweakly.

Conversely, assume
$x_{ij}\to x_i$ ultraweakly in $M$ for all $i\in\{1,\dotsc,n\}$.
Then, for $\sum_{i=1}^n\hat{a}_i\otimes\varphi_i\in A^*\otimes M_*$,
\begin{align*}
\abs[\Big]{\Bigl(\sum_{i=1}^n\hat{a}_i\otimes\varphi_i\Bigr)(z-z_j)}
&= \abs[\Big]{\Bigl(\sum_{i=1}^n\hat{a}_i\otimes\varphi_i\Bigr)
\Bigl(\sum_{i=1}^n a_i\otimes (x_i-x_{ij})\Bigr)} \\
&= \abs[\Big]{\sum_{i=1}^n\varphi_i(x_i-x_{ij})} \\
&\le \sum_{i=1}^n\abs{\varphi_i(x_i)-\varphi_i(x_{ij})}
\to 0
\end{align*}
because $\varphi_i(x_{ij})\to \varphi_i(x_i)$ for all $i$.
Hence $z_j\to z$ ultraweakly in $A\odot M$.
\end{myproof}

\begin{mycorollary}
\label{cor:uwlim-mat-compatible}
Let $M$ be a $W^*$-algebra.
Let $(x_j)_j=([x_{klj}]_{kl})_j$ be a net in $\mat_n(M)$,
and let $x=[x_{kl}]_{kl}\in \mat_n(M)$.
Then, $x_j\to x$ ultraweakly in $\mat_n(M)$ if and only if
$x_{klj}\to x_{kl}$ ultraweakly in $M$ for all $k,l\in\{1,\dotsc,n\}$.
In other words, one has $\uwlim_j [x_{klj}]_{kl} =
[\uwlim_j x_{klj}]_{kl}$, where $\uwlim$ denotes the ultraweak limit.
\qed
\end{mycorollary}

\begin{mylemma}
\label{lem:wtensor-uwcontinuous}
Let $M,N$ be $W^*$-algebra.
Let $x\in M$, and
assume that a norm-bounded net $(y_i)_i$
converges ultraweakly to $y$ in $N$.
Then a net $(x\otimes y_i)_i$
converges ultraweakly to $x\otimes y$
in $M\wtensor N$.
\end{mylemma}
\begin{myproof}
Recall that
\begin{gather*}
y_i \to y\;\;\text{ultraweakly in $N$}
\iff\forall\varphi\in N_*\ldotp
\varphi(y_i)\to\varphi(y)
\\
x\otimes y_i \to x\otimes y\;\;\text{ultraweakly in $M\wtensor N$}
\iff\forall\xi\in (M\wtensor N)_*\ldotp
\xi(y_i)\to\xi(y)
\enspace.
\end{gather*}
Then, for any $\varphi\in M_*$ and $\psi\in N_*$,
\begin{align*}
(\varphi\otimes\psi)(x\otimes y_i)
&=
\varphi(x)\cdot\psi(y_i) \\
&\to
\varphi(x)\cdot\psi(y)
= (\varphi\otimes\psi)(x\otimes y)
\enspace,
\end{align*}
because $\psi(y_i)\to\psi(y)$.
Hence we have $\chi(x\otimes y_i)\to\chi(x\otimes y)$
for all $\chi\in M_*\odot N_*$.
Now, take arbitrary $\xi\in M_*\otimes N_*\cong (M\wtensor N)_*$.
Then, there exists a sequence $(\chi_j)_j$ in $M_*\odot N_*$
convergent to $\xi$ under the dual spatial $C^*$-norm
(therefore, we have an inequation like
$\abs{\xi(z)}\le\norm{\xi}\norm{z}$).
Then
\begin{align*}
&\abs{
\xi(x\otimes y)
- \xi(x\otimes y_i)} \\
&\le
\abs{
\xi(x\otimes y)
- \chi_j(x\otimes y)} +
\abs{
\chi_j(x\otimes y)
- \chi_j(x\otimes y_i)} +
\abs{
\chi_j(x\otimes y_i)
- \xi(x\otimes y_i)} \\
&\le
\norm{\xi-\chi_j}
\norm{x\otimes y} +
\abs{
\chi_j(x\otimes y)
- \chi_j(x\otimes y_i)} +
\norm{\xi-\chi_j}
\norm{x\otimes y_i} \\
&=
\norm{\xi-\chi_j}
\norm{x}\norm{y} +
\abs{
\chi_j(x\otimes y)
- \chi_j(x\otimes y_i)} +
\norm{\xi-\chi_j}
\norm{x}\norm{y_i} \\
&=
\abs{
\chi_j(x\otimes y)
- \chi_j(x\otimes y_i)}
+
\norm{\xi-\chi_j}
\norm{x}(\norm{y}+\norm{y_i})
\enspace.
\end{align*}
Take arbitrary $\varepsilon>0$.
Because $(y_i)_i$ is norm-bounded,
and $\chi_j\to\xi$,
we have
\[
\abs{
\chi_j(x\otimes y)
- \chi_j(x\otimes y_i)}
+
\norm{\xi-\chi_j}
\norm{x}(\norm{y}+\norm{y_i})
<
\abs{\chi_j(x\otimes y)
- \chi_j(x\otimes y_i)}
+\varepsilon
\]
for large enough $j$.
Finally, since $\chi_j(x\otimes y_i)\to
\chi_j(x\otimes y)$, we have
\[
\abs{
\xi(x\otimes y)
- \xi(x\otimes y_i)} <
\abs{\chi_j(x\otimes y)
- \chi_j(x\otimes y_i)}
+\varepsilon
<
2\varepsilon
\]
for sufficiently large $i$.
This proves $\xi(x\otimes y_i)\to\xi(x\otimes y)$.
Hence $x\otimes y_i$
converges ultraweakly to
$x\otimes y$
in $M\wtensor N$.
\end{myproof}

\subsection{Proofs for
$\Dcppobot$-enrichment of $\Wstar_\CPPU$}
\label{subsec:proof-for-enrichment}

\begin{mylemma}
\label{lem:hom-sup-is-uwlim}
The map $f$ defined in the proof of
Proposition~\ref{prop:hom-wstarcppu-is-dcppo}
satisfies $f(x)=\uwlim_i f_i(x)$
for all $x\in M$.
\end{mylemma}
\begin{myproof}
By the definition of $f$ and
by Proposition~\ref{prop:wstar-monotone-complete},
we have $f(x)\coloneqq\sup_i f_i(x)=\uwlim_i f_i(x)$
for $x\in[0,1]_M$.
Because of the linearity
and the ultraweak continuity
of the addition and the multiplication,
we obtain $f(x)=\uwlim_i f_i(x)$ for all $x\in M$.
\end{myproof}

\begin{myproposition}
\label{prop:hom-sup-cp}
The map $f$ defined in the proof of
Proposition~\ref{prop:hom-wstarcppu-is-dcppo}
is completely positive, and
the supremum of $(f_i)_i$.
\end{myproposition}
\begin{myproof}
Consider $\mat_n(f)\colon \mat_n(M)\to\mat_n(N)$.
For positive $[x_{kl}]_{kl}\in\mat_n(M)$,
\begin{align*}
\mat_n(f)([x_{kl}]_{kl})
&=
[f(x_{kl})]_{kl} \\
&=
[\uwlim\nolimits_i f_i(x_{kl})]_{kl}
&& \text{by Lemma~\ref{lem:hom-sup-is-uwlim}}
\\
&=
\uwlim\nolimits_i [f_i(x_{kl})]_{kl}
&& \text{by Corollary~\ref{cor:uwlim-mat-compatible}}
\\
&=
\uwlim\nolimits_i \mat_n(f_i)([x_{kl}]_{kl})
\enspace.
\end{align*}
Now, $f_i$ is completely positive,
so that $\mat_n(f_i)([x_{kl}]_{kl})$
is positive for all $i$.
Moreover, $f_i\sqsubseteq f_j$
implies $\mat_n(f_i)([x_{kl}]_{kl})\le
\mat_n(f_j)([x_{kl}]_{kl})$.
Hence $(\mat_n(f_i)([x_{kl}]_{kl}))_i$
is a positive monotone net in $\mat_n(N)$,
which is bounded because each $f_i$ is subunital
and so is $M_n(f_i)$.
By Proposition~\ref{prop:wstar-monotone-complete}
(and Corollary~\ref{cor:mat-wstar-is-wstar})
we obtain
\[
\mat_n(f)([x_{kl}]_{kl})
=\uwlim\nolimits_i \mat_n(f_i)([x_{kl}]_{kl})
=\sup\nolimits_i \mat_n(f_i)([x_{kl}]_{kl})
\enspace.
\]
Thus $\mat_n(f)([x_{kl}]_{kl})\ge \mat_n(f_i)([x_{kl}]_{kl})\ge 0$,
so that $f$ is completely positive and
$f_i\sqsubseteq f$ for all $i$.
Let $f'\in\Wstar_\CPPU(M,N)$ with $f_i\sqsubseteq f'$
for all $i$.
Then, for positive $[x_{kl}]_{kl}\in\mat_n(M)$,
we have $\mat_n(f_i)([x_{kl}]_{kl})\le \mat_n(f')([x_{kl}]_{kl})$
for all $i$.
Hence $\mat_n(f)([x_{kl}]_{kl})
=\sup\nolimits_i \mat_n(f_i)([x_{kl}]_{kl})
\le \mat_n(f')([x_{kl}]_{kl})$.
It follows that $f\sqsubseteq f'$.
\end{myproof}

\noindent
Note that now Lemma~\ref{lem:hom-sup-is-uwlim}
gives a explicit formula
$(\bigsqcup_i f_i)(x)=\uwlim_i f_i(x)$
for all $x\in M$.

\begin{mylemma}[{\cite[Lemma~3.2.6]{AbramskyJ1994}}]
\label{lem:scott-conti-separate}
Let $P,Q,R$ be posets.
Then a map
$f\colon P\times Q\to R$ is Scott-continuous
if and only if it is separately Scott-continuous.
\qed
\end{mylemma}

\begin{myproposition}
\label{prop:composition-scott-conti}
Let $M,N,L$ be $W^*$-algebras.
The composition
\[
{\circ}\colon
\Wstar_\CPPU(N,L)\times
\Wstar_\CPPU(M,N)\longto
\Wstar_\CPPU(M,L)
\]
is bi-strict Scott-continuous.
\end{myproposition}
\begin{myproof}
The bi-strictness is obvious because
bottom maps $\bot$ are the zero maps.
By Lemma~\ref{lem:scott-conti-separate},
it suffices to show the Scott-continuity in each variable.
Let $(g_i)_i$ be a monotone net in $\Wstar_\CPPU(N,L)$
and let $f\in\Wstar_\CPPU(M,N)$.
It is easy to see $(-)\circ f$ is monotone,
and hence $(g_i\circ f)_i$ is a monotone net in
$\Wstar_\CPPU(M,L)$. Then for each $x\in[0,1]_M$,
\[
\bigl(\bigl(\bigsqcup\nolimits_i g_i\bigr)\circ f\bigr)(x)
= \bigl(\bigsqcup\nolimits_i g_i\bigr)(f(x))
= \sup\nolimits_i g_i(f(x))
= \sup\nolimits_i (g_i\circ f)(x)
= \bigl(\bigsqcup\nolimits_i (g_i\circ f)\bigr)(x)
\]
so that $(\bigsqcup_i g_i)\circ f= \bigsqcup_i (g_i\circ f)$.

Let $(f_i)_i$ be a monotone net in $\Wstar_\CPPU(M,N)$
and let $g\in\Wstar_\CPPU(N,L)$.
It is easy to see $g\circ (-)$ is monotone,
and hence $(g\circ f_i)_i$ is a monotone net in
$\Wstar_\CPPU(M,L)$.
Then for each $x\in[0,1]_M$,
\[
\bigl(g\circ \bigl(\bigsqcup\nolimits_i f_i\bigr)\bigr)(x)
= g(\sup\nolimits_i f_i(x))
= \sup\nolimits_i g(f_i(x))
= \sup\nolimits_i (g\circ f_i)(x)
= \bigl(\bigsqcup\nolimits_i (g\circ f_i)\bigr)(x)
\enspace,
\]
where we used the normality of $g$ (and
Proposition~\ref{prop:normal-sup-pres-prelim}) for the second equality.
It shows $g\circ (\bigsqcup_i f_i)
= \bigsqcup\nolimits_i (g\circ f_i)$.
\end{myproof}

\begin{myproposition}
\label{prop:tupling-scott-conti}
Let $M,N,L$ be $W^*$-algebras.
Then the tupling of maps
\[
\tuple{\cdot,\cdot}\colon
\Wstar_\CPPU(L,M)
\times
\Wstar_\CPPU(L,N)
\stackrel{\cong}{\longto}
\Wstar_\CPPU(L,M\oplus N)
\]
is strict Scott-continuous.
\end{myproposition}
\begin{myproof}
The strictness is easy.
Because of the symmetry, and by Lemma~\ref{lem:scott-conti-separate},
it suffices to show the Scott-continuity in the first variable.
Let $(f_i)_i$ be a monotone net in $\Wstar_\CPPU(L,M)$
and let $g\in\Wstar_\CPPU(L,N)$.
It is easy to see $\tuple{\cdot,g}$ is monotone,
and hence $(\tuple{f_i,g})_i$ is a monotone net in
$\Wstar_\CPPU(L,M\oplus N)$.
Then for each $x\in [0,1]_L$,
\begin{align*}
\bigl(\bigsqcup\nolimits_i \tuple{f_i,g}\bigr)(x)
&= \sup\nolimits_i \tuple{f_i,g}(x) \\
&= \sup\nolimits_i (f_i(x),g(x)) \\
&= (\sup\nolimits_i f_i(x),\sup\nolimits_i g(x)) \\
&= \Bigl(\bigl(\bigsqcup\nolimits_i f_i\bigr)(x),g(x)\Bigr) \\
&= \tuple[\big]{\bigsqcup\nolimits_i f_i,g}(x)
\enspace.
\end{align*}
Note that the order in $M\oplus N$ is pointwise,
therefore the supremum is pointwise, too.
Hence $\bigsqcup_i \tuple{f_i,g}= \tuple{\bigsqcup_i f_i,g}$.
\end{myproof}

\begin{myproposition}
\label{prop:wtensor-scott-conti}
Let $M,M',N,N'$ be $W^*$-algebras.
The tensor product of maps
\[
{\wtensor}\colon
\Wstar_\CPPU(M,M')\times
\Wstar_\CPPU(N,N')
\longto
\Wstar_\CPPU(M\wtensor N,M'\wtensor N')
\]
is bi-strict Scott-continuous.
\end{myproposition}
\begin{myproof}
Let $f\in \Wstar_\CPPU(M,M')$.
By the symmetry and Lemma~\ref{lem:scott-conti-separate},
it suffices to show
\[
f\wtensor(-)\colon
\Wstar_\CPPU(N,N')
\to\Wstar_\CPPU(M\wtensor N,M'\wtensor N')
\]
is strict Scott-continuous.
Let $\bot\in \Wstar_\CPPU(N,N')$ be
the least element, i.e.\ the zero map.
Then
\[(f\wtensor \bot)(x\otimes y) =
f(x)\otimes \bot(y) = f(x)\otimes 0 = 0
\]
for all $x\in M,y\in N$.
Hence $(f\wtensor \bot)(z)=0$ for all $z\in M\odot N$.
Because $M\odot N$ is ultraweakly dense in $M\wtensor N$
and $f\wtensor \bot$ is normal (i.e.\ ultraweakly continuous),
we obtain $(f\wtensor \bot)(z)=0$ for all $z\in M\wtensor N$.
Therefore $f\wtensor \bot=\bot$.

Let $g,g'\in \Wstar_\CPPU(N,N')$
with $g\sqsubseteq g'$. By definition $g'-g$ is completely positive,
and so is $f\wtensor (g'-g)$.
Notice that $f\wtensor (g'-g)=f\wtensor g'-f\wtensor g$
because they coincide on $M\odot N$.
Then $f\wtensor g'-f\wtensor g$ is completely positive, and
$f\wtensor g\sqsubseteq f\wtensor g'$.
Hence $f\wtensor(-)$ is monotone.

Let $(g_i)_i$ be a monotone net in $\Wstar_\CPPU(N,N')$.
By the monotonicity $(f\wtensor g_i)_i$ is a monotone net in
$\Wstar_\CPPU(M\wtensor N,M'\wtensor N')$.
By a similar argument above,
to prove $f\wtensor(\bigsqcup_i g_i)=\bigsqcup_i(f\wtensor g_i)$,
it suffice to show
$(f\wtensor(\bigsqcup_i g_i))(x\otimes y)=
(\bigsqcup_i f\wtensor g_i)(x\otimes y)$
for all $x\in M,y\in N$.
This is showed as follows.
\begin{align*}
\bigl(f\wtensor\bigl(\bigsqcup\nolimits_i g_i\bigr)\bigr)(x\otimes y)
&= f(x)\wtensor\bigl(\bigsqcup\nolimits_i g_i\bigr)(y) \\
&= f(x)\otimes(\uwlim\nolimits_i g_i(y))
&&\text{by Lemma~\ref{lem:hom-sup-is-uwlim}}\\
&= \uwlim\nolimits_i (f(x)\otimes g_i(y))
&&\text{by Lemma~\ref{lem:wtensor-uwcontinuous}}\\
&= \uwlim\nolimits_i(f\wtensor g_i)(x\otimes y) \\
&= \bigl(\bigsqcup\nolimits_i f\wtensor g_i\bigr)(x\otimes y)
&&\text{by Lemma~\ref{lem:hom-sup-is-uwlim}}
\end{align*}
\end{myproof}

\end{document}